\documentclass[fleqn,usenatbib]{mnras}

\usepackage{graphicx}
\usepackage{amssymb}
\usepackage{amsmath}
\usepackage{mathabx}
\usepackage{txfonts}
\usepackage[T1]{fontenc}
\usepackage{ae,aecompl}
\usepackage{cleveref}
\usepackage{comment}
\usepackage{algorithm2e}
\usepackage{comment}

\title[SD method for the Euler equations]
{Spectral Difference method with {\it a posteriori} limiting: Application to the Euler equations in one and two space dimensions}

\author[D. A. Velasco Romero et al]{%
David A. Velasco Romero$^{1,3}$\thanks{E-mail: david.velasco@ics.uzh.ch} , Maria Han Veiga$^{2}$ , Romain Teyssier$^{1,3}$\\
$^{1}$Department of Astrophysical Sciences, Princeton University,
4 Ivy Lane,
 Princeton, New Jersey 08544, United States.\\
$^{2}$Department of Mathematics and Michigan Institute for Data Science, University of Michigan, 530 Church St, 48109, Ann Arbor, MI,\\ United States of America\\
$^{3}$Institute for Computational Science, University of Zurich, Winterthurerstrasse 190, 8057 Zurich, Switzerland
}

\date{Accepted XXX. Received YYY; in original form ZZZ}

\pubyear{2022}

\begin{document}
\label{firstpage}
\pagerange{\pageref{firstpage}--\pageref{lastpage}}
\maketitle

\begin{abstract}
We present a new numerical scheme which combines the Spectral Difference (SD) method up to arbitrary high order with \emph{a-posteriori} limiting using the classical MUSCL-Hancock scheme as fallback scheme. It delivers very accurate solutions in smooth regions of the flow, while capturing sharp discontinuities without spurious oscillations. We exploit the strict equivalence between the SD scheme and a Finite-Volume (FV) scheme based on the SD control volumes to enable a straightforward limiting strategy. At the end of each stage of our high-order time-integration ADER scheme, we check if the high-order solution is admissible under a number of numerical and physical criteria. If not, we replace the high-order fluxes of the troubled cells by fluxes from our robust second-order MUSCL fallback scheme. We apply our method to a suite of test problems for the 1D and 2D Euler equations. We demonstrate that this combination of SD and ADER provides a virtually arbitrary high order of accuracy, while at the same time preserving good sub-element shock capturing capabilities.
\end{abstract}




\section{Introduction}

Most, if not all, recent advances in theoretical astrophysics are built upon the extensive use
of accurate fluid solvers for the ideal Euler equations.  Many of the popular codes used 
in computational astrophysics, (e.g.: \texttt{RAMSES}, \texttt{FARGO3D}, \texttt{ATHENA++}, \texttt{Pencil}, \texttt{FLASH},  \texttt{GADGET}, \texttt{AREPO}, among others) are based on the finite difference (FD) or finite volume (FV) framework, with an emphasis on properly capturing strong shock waves. Comparatively, finite element (FE) codes are much less used in astrophysics, although they are now routinely used in e.g. engineering and geophysics. 

A key aspect of computational astrophysical fluid dynamics is the ability to deliver higher and higher accuracy for longer and longer time integration. Modern architectures such as massively parallel computers or graphical processing units (GPU) can be leveraged to increase the spatial and time resolution of the simulations. This requires the algorithms to be adapted to this new hardware and to overcome several issues related to data locality and communication cost. 

Another strategy is to increase the order of accuracy of the schemes. Indeed, for a given target accuracy, higher order methods are well known to be more efficient for smooth solutions of the Euler equations.
In that context, FE methods have gained popularity in the computational astrophysics community. First, FE methods provide an elegant framework to build code of higher and higher order of accuracy, while requiring less computations and less communication than their corresponding high-order FD or FV schemes. Second, FE methods are very computationally intensive with good data locality, making them ideally suited for GPUs. 

Within the broader FE family, the Discontinuous Galerkin (DG) method \citep{reed1973triangular,cockburn1989tvb, cockburn1991runge, cockburn1998} has emerged as a particularly interesting method for astrophysical applications of the Euler equations in the last decade \citep{Schaal2015,kidder2017spectre,Velasco2018}. Several successful attempts have been made to also develop DG for the ideal MHD equations \citep{Mocz2014, zanotti2015solving, anninos2017cosmosdg,fambri2018ader,Guillet2019,fambri2020discontinuous}. The DG method is particularly interesting because calls to the Riemann solver are only required at the boundary of the elements, with however a rather restrictive Courant stability condition with $\Delta t < h/S/(2p+1)$, where $h$ is the size of an element, $S$ is the maximum wave speed, $p$ is the polynomial degree of the DG method and $p+1$ is the corresponding order of accuracy.

One of the most challenging aspects discussed in this large body of work is for DG to be able to capture properly sharp discontinuities without triggering spurious oscillations. This can be achieved using artificial viscosity techniques \citep[e.g][]{richtmyer1950method,jameson2014,Lu2019} where a dissipation mechanism is added explicitly, either everywhere in the flow or only within the shock region. 

Another strategy consists in enforcing the monotonicity of the numerical solution using limiters \citep{1979JCoPh..32..101V,sweby1984high,qiu2005comparison,krivodonova2007limiters,may2012}. 

Each strategy has pros and cons. The artificial viscosity approach effectively modifies the ideal fluid into a viscous fluid, while limiters degrade the order of the method locally down to first order, throwing away most of the additional internal degrees of freedom (DOF) within the element. Indeed, the DG method with polynomial degree $p$ has $p+1$ DOF per element in 1D. If a discontinuity is detected nearby, most limiters will just reset all the coefficients corresponding to high degree polynomials to zero, only keeping the average, effectively falling back to a piecewise constant approximation of the solution within the element. The corresponding FV scheme would have $p+1$ cells per element, with a less restrictive Courant condition $\Delta t <h/S/(p+1)$ but one call to the Riemann solver for each cell face.

A new approach has been proposed recently by \cite{Dumbser2014} to overcome this limitation of DG by overlaying a small Cartesian grid on top of these {\it troubled} elements, preserving the internal DOF using a monotonicity preserving fallback scheme such as the MUSCL\footnote{Monotonic Upstream-centered Scheme for Conservation Laws}  scheme introduced by \citet{1979JCoPh..32..101V}, the PPM\footnote{Piecewise Parabolic Method} scheme introduced by \citet{colella1984piecewise}, the ENO\footnote{Essentially Non-Oscillatory} scheme introduced by \citet{harten1987uniformly}, or the WENO\footnote{Weighted Essentially Non-Oscillatory} scheme introduced by \citet{liu1994weighted}. 

Though all these methods differ by construction, they all provide a robust monotonicity preserving solution suppressing  oscillations in presence of discontinuities. Since the trouble element detection is performed after the unlimited  DG step, these techniques are usually referred to as {\it a posteriori limiting}. The small Cartesian grid introduced locally to replace the DG solution contains $2p+1$ cell per dimension so that the Courant stability condition of the FV fallback scheme matches exactly the one of the original DG scheme. 

As there exist an equivalence between FE and FV methods \citep{Dumbser2014}, there is the possibility of pairing a high-order FE method with a bounded FV method as the fallback scheme \citep{Dumbser2014, Vilar2019}. There exist already several different implementations of this approach \citep{zanotti2015space, bacigaluppi2019a, Vilar2019}. In that context, the Multi-dimensional Optimal Order Detection (MOOD) technique as been proposed as a general framework for applying these ideas to almost every type of high-order methods \citep{clain2011high, diot2012improved, diot2013multidimensional, loubere2014new}. 

This method can be however quite complex to implement in practice due to the fundamentally hybrid nature of the data representation, switching all the time between the DG and the FV solution vectors. Sub-cell (or sub-element) shock capturing techniques have been proposed to overcome this limitation \citep{Huerta2012, Sonntag2014, markert2021}. The idea relies heavily on the strict equivalence between the so-called quadrature-free nodal DG approach and its corresponding FV discretization. It is possible to use the exact same data representation for both the DG method and the fallback FV scheme that preserves the monotonicity of the solution. 

In this work, instead of the classical DG scheme, we used the Spectral Difference (SD) method \citep{Liu2006} which can be shown to be equivalent to a quadrature free nodal DG scheme \citep{May2011}. The SD method is appealing because it only requires (like DG) one call to the Riemann solver at the edge of the elements, not inside the elements. The less stringent Courant stability condition is however $\Delta t <h/S/(p+1)$, the same as the corresponding FV scheme.
The SD method can provide, like DG, an arbitrarily high-order approximation in space, with however a natural interpretation as a FV scheme without the need to hybridize the data representation, a nice property that we exploit in this paper. 

As it is desirable to pair the spatial order of accuracy provided by these high-order methods with an equivalent temporal order of accuracy, a lot of work has been dedicated into the development of high-order time integrators. Strong Stability Preserving (SSP) Runge-Kutta (RK) is a widely used family of implicit and explicit high-order time integrators \citep[for a recent review see][]{gottlieb2009high}. More recently, the DG approach has been also implemented in a spatial-temporal discretization \citep{Dumbser2008,balsara2009,dumbser_ader_2013}, giving rise to a new flavor of time integrators called ADER method, first introduced in the FV context by \citet{toro2001towards}, and later extended by \citet{titarev2002ader,titarev2005ader}. When coupled to DG or SD, this approach results in a generic method capable of providing  arbitrarily high order for both spatial and temporal discretization. We will adopt the ADER method in this paper.

We explore in this paper a novel SD-ADER scheme with sub-cell shock capturing capabilities thanks to a posteriori limiting using a second-order monotonicity preserving fallback scheme. We apply this new method to both the advection equation and the Euler equations in 1 and 2 space dimensions. The paper is divided as follows: in \autoref{sec:method} we start by presenting a summary of the SD method and the ADER scheme. In \autoref{sec:limiting} we present the \textit{a posteriori} limiting methodology. We then present in \autoref{sec:results} a series of tests for both the advection equation and the Euler equations in 1D and 2D. Finally, in \autoref{sec:discussion} we present a discussion of the results, and we close in \autoref{sec:conclusions} with our conclusions.

\section{Numerical method}
\label{sec:method}

\subsection{Spectral Differences}

We start by presenting the SD method for a one dimensional scalar conservation law:
\begin{equation}
\partial_t u + \partial_x f( u ) = 0,
\label{eq:1}
\end{equation}
defined in a spatial domain $\Omega$. We partition the domain with non-overlapping intervals (elements). 
The continuous high-order numerical solution $u(x)$ inside element $a$ is given by:
\begin{equation}
    u(x) = \sum_{m=0}^p u(x^s_{a,m})\ell^s_m(x),
    \label{eq:2}
\end{equation}
where $\{\ell^s_{m}(x)\}_{m=0}^p$ represents the set of Lagrange interpolation polynomials up to degree $p$ built on the set of $p+1$ points $\mathcal{S}^s=\{x^s_{a,m}\}_{m=0}^p$ in element $a$, called the solution points (hence the superscript $s$). 

The discrete numerical solution is defined as the value of the solution at these solution points, namely:
\begin{equation} 
u^s_{a,m}=u(x^s_{a,m}).
\end{equation}
and the solution vector is defined as 
\begin{equation}
{\bf u} = \left\{u^s_{a,m}\right\}~{\rm with}~a=1,N~{\rm and}~m=0,p .
\end{equation}

Using this Lagrange polynomial, we compute the solution at the set of $p+2$ points $\mathcal{S}^f=\{x^f_{a,m}\}_{m=0}^{p+1}$ in element $a$, called the flux points (hence the superscript $f$). The solution at flux point is defined by:
\begin{equation} 
u^f_{a,m}=u(x^f_{a,m}).
\end{equation}
Note that the first flux point $x^f_{a,0}$ is located exactly at the left boundary of element $a$, while the last flux point $x^f_{a,p+1}$ is located exactly at the right boundary of element $a$.

The high-order approximation of the flux is given by:
\begin{equation}
f(x) =  \sum_{m=0}^{p+1} f(u^f_{a,m}) \ell^f_m(x),
\label{eq:3}
\end{equation}
where now $\{\ell^f_{m}(x)\}_{m=0}^{p+1}$ is another set of Lagrange interpolation polynomials of up to degree $p+1$ built on the set of $p+2$ points flux points. 

\begin{figure}
    \centering
    \includegraphics[width=\columnwidth]{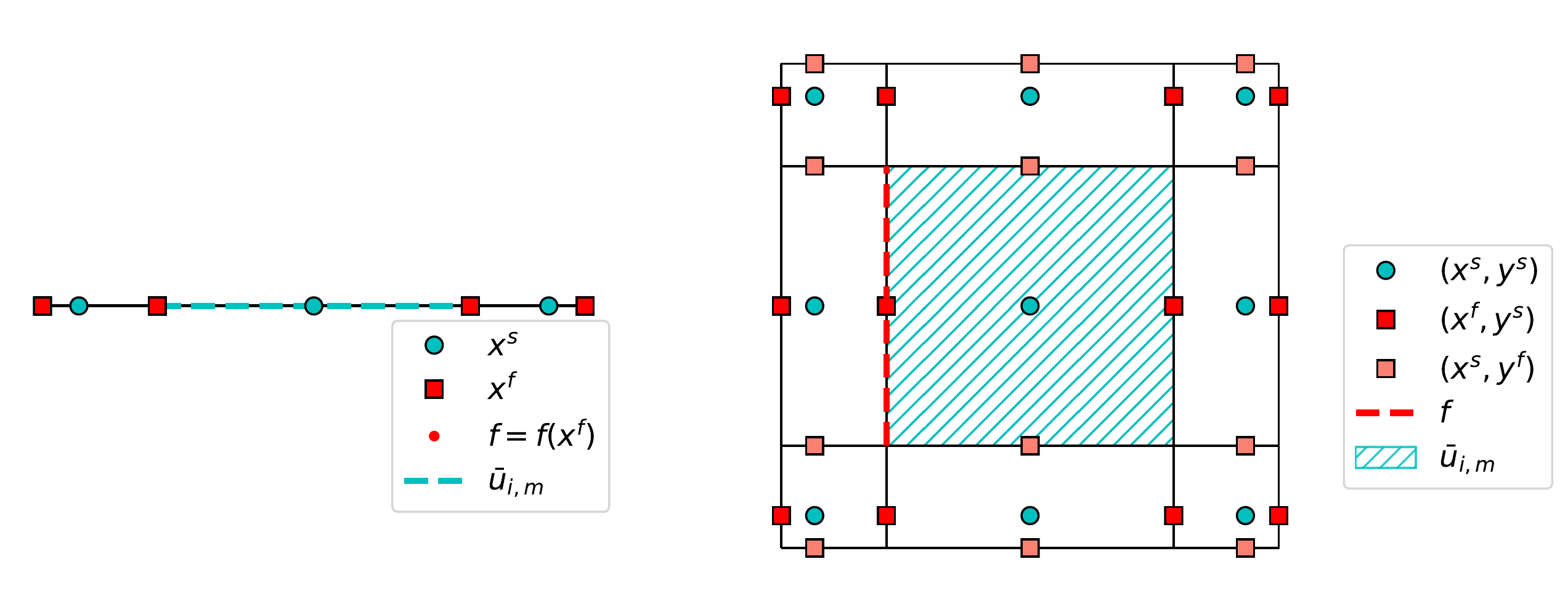}
    \caption{Visualization of our Cartesian finite element for $p=2$. Circles denote the solution points while flux points are represented as squares.}
    \label{fig:1}
\end{figure}

\begin{figure}
    \centering
    \includegraphics[width=\columnwidth]{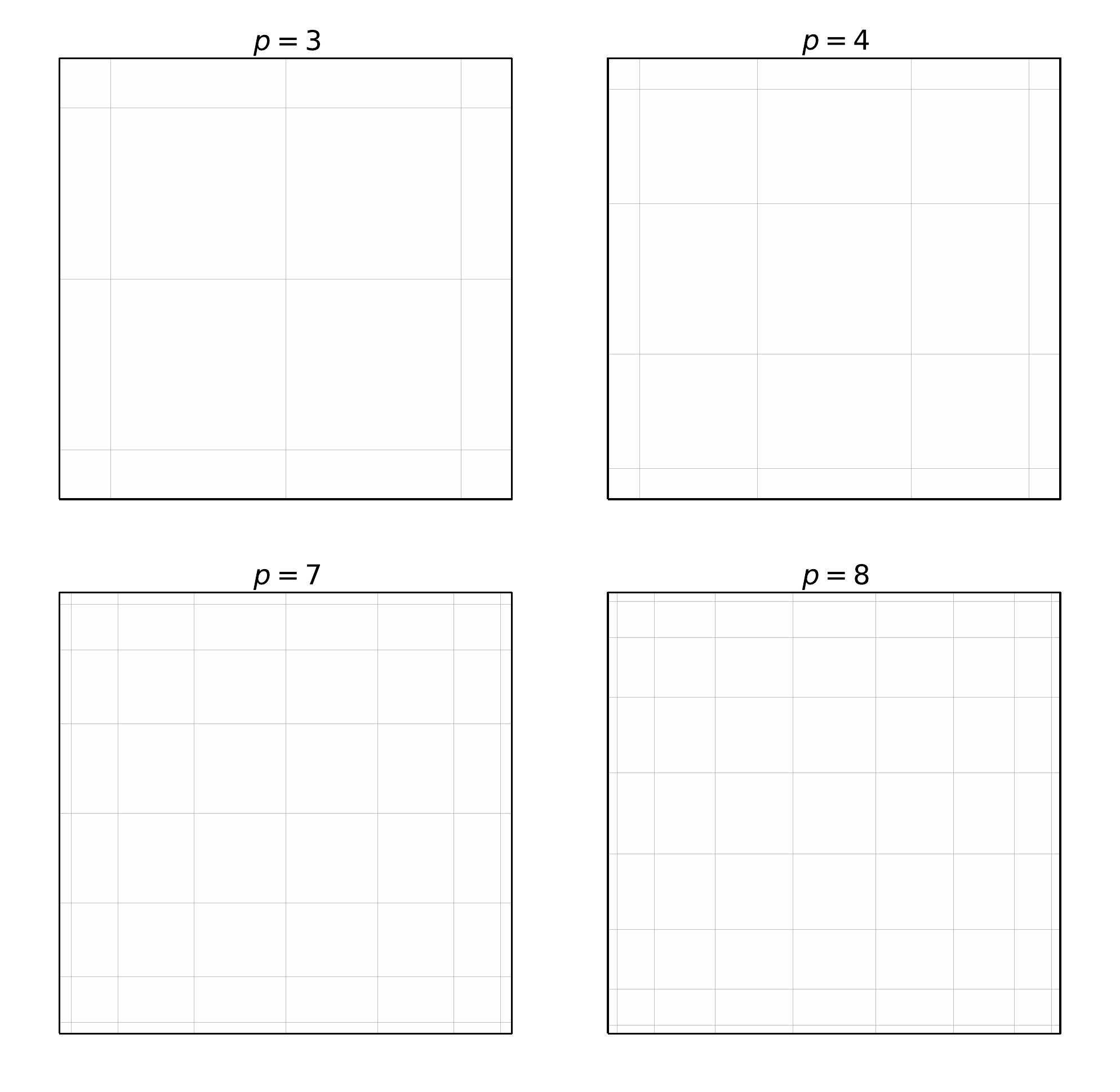}
    \caption{Visualization of the control volumes for different polynomial degrees. For stability at higher and higher polynomial degree, the flux points have to be chosen closer and closer to the element boundaries.}
    \label{fig:2}
\end{figure}

At this stage, the flux at the boundary of the elements is discontinuous. We compute a continuous flux function by solving a Riemann problem at the interface between elements. Let us denote $\hat{f}(\cdot)$  the numerical flux resulting from solving this Riemann problem. 
The update of the solution $u(x,t)$ can be now obtained through the exact derivative   
of the flux evaluated at the solution points:
\begin{equation}
\hat{f}'(x^s_m) =  \sum_{m'=0}^{p+1} \hat{f}(u^f_{a,m'}) \ell'^f_{m'}(x^s_m),
\end{equation}
where $\ell^{\prime}$ is the derivative of the Lagrange polynomials.
This results in the following semi-discrete scheme:
\begin{align}
\label{eq:semi-discrete}
\frac{{\rm d}}{{\rm d}t} u^s_{a,m}(t) = {\cal L}_{a,m}(\bold{u})= -\sum_{m'=0}^{p+1} \hat{f}(u^f_{a,m'}(t)) \ell^{\prime f}_{m'}(x^s_{a,m}).
\end{align}

In \cite{jameson2010}, the one-dimensional SD method is proven to be stable at all orders, 
provided the interior flux points are placed at the zeros of the corresponding Legendre polynomials (Gauss-Legendre quadrature points).  
In \cite{abeele2008}, the stability of the SD method for two-dimensional Cartesian meshes has also been proven. As long as solution points are contained anywhere within the control volume delimited by the flux points, the stability of the method depends only on the position of the flux points. As such, we use Gauss-Legendre quadrature points for the inner flux points and the zeros of the Chebyshev polynomials for the solution points. 

The corresponding solution points and flux points are shown in Figure~\ref{fig:1} for a typical square element and for $p=2$. We see that the end flux points along each dimension are sitting exactly at the element boundaries. Flux points in each dimension also define small subcells inside each element that we call here {\it control volumes}. \autoref{fig:2} presents elements for $p=3,4,7$ and $8$, showing their inner control volumes. These control volumes play a fundamental role in our numerical scheme.

\subsection{Time integration}

For the time integration, we use the ADER scheme presented in \citet{Dumbser2008,balsara2009,mhv2020}. 
In this method, the solution of a general ODE of the form
\begin{equation}
\frac{d}{dt}u(t) = {\cal L}(u(t))
\end{equation}
is represented using Lagrange polynomials {\it in time} $\ell_k(t)$ defined using the $p+1$ Gauss-Legendre quadrature points 
$\lbrace t_k \rbrace_{k=0}^p \in [0,\Delta t]$,  with corresponding quadrature weights $\lbrace w_k \rbrace_{k=0}^p$:
\begin{equation}
u(t) = \sum_{k=0}^p u_{k}\ell_k(t), 
\end{equation}
where $p$ is equal to the polynomial degree of the spatial discretization, as we aim for a solution with the same order of accuracy in space and time. 
As described in \citet{HANVEIGA2021110327}, this method can be written in a compact form using the mass matrix $M \in \mathbb{R}^{(p+1)\times(p+1)}$ and a vector $r$:
\begin{equation}\label{fix:point}
 \sum_{k'=0}^{p}M_{kk'} u_{k'} = r_k(u_0,...,u_p).
\end{equation}
This system is then solved by means of a Picard iterative scheme:
\begin{equation}
\label{eq:fixpoint_iteration}
u^{q+1}_k=\sum_{k=0}^p M^{-1}_{kk'} r_{k'} (u^{q}_0,...,u^{q}_p),
\end{equation}
where the superscript $q$ represents the iteration step. This implicit method requires of an initial guess. 
Here we use $\lbrace u^{0}_k=u(t)\rbrace_{k=0}^p$, the solution at the end of the previous time step. 
Finally, we use these final predicted states $\lbrace u^{p}_k \rbrace_{k=0}^p$ at our time quadrature points to obtain the final solution:
\begin{equation}
u(t + \Delta t) = u(t) + \Delta t \sum_{k=0}^p w_{k} {\cal L}(u^{p}_k).
\label{eq:sd_update}
\end{equation}
This method only requires $p$ internal corrections to the solution (Picard iterations) to obtain a solution that is accurate up to order $p+1$ in time \citep{dumbser_ader_2013}. As such, the first order scheme ($p=0$) (no iteration) corresponds exactly to the first-order forward Euler scheme. However, contrary to  \cite{dumbser_ader_2013}, we solve the Riemann problem at element boundaries and at each time slice.

The Courant condition for stability of the SD scheme \citep{vanharen2017} is:
\begin{equation}
\Delta t = \frac{C}{p+1} \frac{h}{|v_{\rm max}|},    
\end{equation}
where $C<0.8-1$ is the Courant factor and $p$ is the polynomial degree of our discretization in space and in time. This choice of Courant factor is justified in section~3 of \citet{HANVEIGA2021110327}.

\subsection{Equivalence to Finite Volume}
\label{sec:sdisfv}

In this section we will show a particularly useful property of the SD method, namely its strict equivalence to a FV method. To demonstrate this, let us once again use the simple one-dimensional scalar conservation law given by \autoref{eq:1}. 
Let's define subcell (or control volume) $m$ in element $a$ by the interval $I_{a,m}= x_a + [x^f_{m},x^f_{m+1}]$. We proceed by integrating \autoref{eq:1} in this control volume $I_{a,m}$:
\begin{equation}
  \int_{I_{a,m}}  \frac{du}{dt} dx = -\int_{I_{a,m}}\frac{\partial f(u)}{\partial x} dx,  
  \label{eq:sub}
\end{equation}
we then define the control volume average as: 
\begin{equation}
  \bar{u}_{a,m} = \frac{1}{h_m}\int_{I_{a,m}} {u}(x) dx.
  \label{eq:cv}
\end{equation}
where $h_m=x^f_{m+1}-x^f_m$ is the size of the corresponding control volume. Inserting \autoref{eq:2} in the right side of \autoref{eq:cv} the control volume average reads:
\begin{equation}
    \bar{u}_{a,m} = \sum_{m'}^p u^s_{a,m'} A_{mm'},
\end{equation}
where the matrix $[A_{mm'}] \in \mathbb{R}^{(p+1)\times(p+1)}$ is given by:
\begin{equation}
    A_{mm'} = \frac{1}{h_m}\int_{I_a,m} \ell^s_{m'}(x)dx.
\end{equation}
Replacing \autoref{eq:cv} in \autoref{eq:sub} and making use of the divergence theorem on the right hand side of \autoref{eq:sub}, we obtain:
\begin{equation}
    \label{eq:fv_dudt}
    \frac{d\bar{u}_{a,m}}{dt} = -\frac{{f}^f_{a,m+1}-{f}^f_{a,m}}{h_{m}}. 
\end{equation}
In the one-dimensional case, these fluxes are strictly equivalent to the fluxes at the flux points \autoref{eq:3}:
\begin{equation}
  {f}^{f}_{a,m} = f(u^f_{a,m}),
\end{equation}
as depicted on the left of \autoref{fig:1}.
In the two-dimensional case these fluxes are equivalent to the line integral along the perpendicular direction, as shown by the dashed red line on the right of \autoref{fig:1}.
At the interface between elements, we need a Riemann solver  to obtain a single valued flux function there:
\begin{equation}
    \hat{f}^f_{a,m} = 
    \begin{cases}
    RP\{u^f_{a-1,p+1},u^f_{a,0}\} \quad &m=0\\
    f^f_{a,m} \quad &1\leq m\leq p \\
    RP\{u^f_{a,p+1},u^f_{a+1,0}\} \quad &m=p+1
    \end{cases}
\end{equation}
Injecting these single value fluxes in \autoref{eq:fv_dudt} we obtain the semi-discrete scheme:
\begin{equation}
    \label{eq:fv_dudt_2}
    \frac{d\bar{u}_{a,m}}{dt} = -\frac{\hat{f}^f_{a,m+1}-\hat{f}^f_{a,m}}{h_{m}}. 
\end{equation}
Integrating the semi-discrete scheme in \autoref{eq:semi-discrete} over the control volume $I_{a,m}$, and dividing by its size we retrieve:
\begin{align}
        \frac{1}{h_{m}}\int_{I_{a,m}} \frac{{\rm d}}{{\rm d}t} u^s_{a,m}(t) &= \frac{1}{h_{m}}\int_{I_{a,m}} {\cal L}_{a,m}(\bold{u}) \\ &= -\frac{1}{h_{m}}\int_{I_{a,m}}\hat{f}'(x)dx = -\frac{\hat{f}^f_{a,m+1}-\hat{f}^f_{a,m}}{h_{m}},
\end{align}
which is the semi-discrete scheme of \autoref{eq:fv_dudt_2}.
Therefore, the time-integration in \autoref{eq:sd_update} can be performed equivalently using this FV update, resulting in the fully-discrete scheme:
\begin{equation}
\bar{u}_{a,m}(t + \Delta t) = \bar{u}_{a,m}(t) - \Delta t \sum_{k=0}^p w_k\frac{\left(\hat{f}^{f,k}_{a,m+1}-\hat{f}^{f,k}_{a,m}\right)}{h_{m}}.
\end{equation}

\section{A posteriori subcell limiting}
\label{sec:limiting}

A typical problem of high-order methods is the so called\textit{ Gibbs phenomenon}. In response to the presence of a discontinuity in the solution, large unphysical oscillations appear. When solving nonlinear conservation laws, this problem is particularly severe as these oscillations can potentially lead to negative densities and pressures. Most of the effort in developing high order methods is therefore spent in designing monotonicity and positivity preserving schemes. 

In this work, we follow the strategy called \textit{a posteriori subcell limiting}, as described by \citet{loubere2014new, Vilar2019, Dumbser2018}. This general methodology consists of correcting the high-order fluxes that are responsible for the oscillatory behavior after the time integration has been completed. For sake of simplicity and readability, we will change our notation slightly, introducing the index $i = m + a(p+1)$ for the control volume (subcell) $m$ of element $a$. With this notation, each subcell is updated using the FV formulation of SD as follows:
\begin{equation}
    \bar{u}^{k+1}_{i} = \bar{u}^{k}_{i}  -w_k\frac{\left(\hat{f}^{f,k}_{i+1/2}-\hat{f}^{f,k}_{i-1/2}\right)}{h_{i}}\Delta t.
\end{equation}

After each stage $k$ of the ADER time integration, the resulting new solution $\bar{u}^{k+1}_{i}$ has to comply with a series of tests based on several numerical and physical requirements. If these requirements are violated, the cell is marked as a \textit{troubled cell} and all high-order fluxes modifying the conservative variables of this subcell are replaced by lower order fluxes using a monotonicity and positivity preserving method, called here \textit{the fallback scheme}.

\subsection{Fallback scheme}

In this paper, we use a classical second order Godunov scheme \citep{1979JCoPh..32..101V} as the fallback scheme used to recompute fluxes of the troubled subcells. This method is based on a piecewise linear approximation for the solution:
\begin{equation}
    u_i(x) = \bar{u}_i + (x-\bar{x}_i)S_i, \quad x \in [x^f_{i-1/2},x^f_{i+1/2}]
\end{equation}
where $\bar{u}_i$ is again the control-volume average, $\bar{x}_i$ is the control-volume centroid, and $S_i = (\partial_x u)_i$ is an estimate of the first derivative of the solution (usually called the slope) at the given subcell. 
At the boundaries of the subcell we reconstruct the solution using:
\begin{equation}
    u_{i-1/2} = \bar{u}_i - \frac{h_i}{2}S_i, \quad u_{i+1/2} = \bar{u}_i + \frac{h_i}{2}S_i,
\end{equation}
where $h_i=x^f_{i+1/2}-x^f_{i-1/2}$.
If we choose the centered second order finite difference approximation of the derivative
\begin{equation}
S_i = \frac{\bar{u}_{i+1}-\bar{u}_{i-1}}{\bar{x}_{i+1}-\bar{x}_{i-1}},
\end{equation}
we end up with a non monotone scheme, as per Godunov's order barrier theorem. However, a monotonicity preserving  scheme can be designed using a slope limiter \citep{harten1983}.

\subsubsection{Slope limiters}

The limited slope is defined by
\begin{equation}
\label{eq:limited_slope}
\tilde{S}_i = lim(\bar{u}_{i-1},\bar{u}_{i},\bar{u}_{i+1})\frac{\bar{u}_{i+1}-\bar{u}_{i-1}}{\bar{x}_{i+1}-\bar{x}_{i-1}},
\end{equation}
where the limiter function is a non linear function of the control volume averaged solutions of the left, middle and right cells that satisfies:
\begin{equation}
    0 \leq lim(\bar{u}_{i-1},\bar{u}_{i},\bar{u}_{i+1}) \leq 1.
\end{equation}

\subsubsection{The \textit{minmod} limiter}
\label{sec:minmod}

Let us define the left and right slopes of subcell $i$ as:
\begin{equation}
S^L_i = \frac{\bar{u}_i - \bar{u}_{i-1}}{\bar{x}_i - \bar{x}_{i-1}}, \quad S^R_i = \frac{\bar{u}_{i+1}-\bar{u}_{i}}{\bar{x}_{i+1}-\bar{x}_{i}},
\end{equation}
The so called \textit{minmod} limiter is defined as:
\begin{equation}
    \tilde{S}_i = \text{sgn}\left(S_i\right) \cdot min(|S^L_i|,|S^R_i|),
\end{equation}
and ensures strict monotonicity of the reconstructed solution.

\subsubsection{The \textit{moncen} limiter}
\label{sec:moncen}

The \textit{moncen} limiter \citep{van1977towards} is slightly less restrictive in the sense that it only enforce a maximum principle of the reconstructed values with respect to the 3 cell-averaged reference values.  We first define the central slope as the average of the left and right slopes:
\begin{equation}
    S^C_i = \frac{1}{2}(S^L_i + S^R_i),
\end{equation}
We then impose the following conditions:
\begin{align*}
\text{If}& \quad S_i>0:\\
&u_{i-\frac{1}{2}} \geq \bar{u}_{i-1},  \quad \to S_i \geq \frac{2}{h_i}(\bar{u}_i -\bar{u}_{i-1}) = \frac{2h^L_{i}}{h_i} S^L_i,\\
&u_{i+\frac{1}{2}} \leq \bar{u}_{i+1},  \quad \to S_i \leq \frac{2}{h_i}(\bar{u}_{i+1} -\bar{u}_{i}) = \frac{2h^R_{i}}{h_i} S^R_i,
\end{align*}
where $h^L_{i} = x_{i}-x_{i-1}$ and $h^R_{i} = x_{i+1}-x_{i}$. The opposite conditions are imposed for $S^C_i < 0$. This leads to the following formula for the limited slope: 
\begin{equation}
    \tilde{S}_i = \text{sgn}\left(S^C_i\right) \cdot \min\left(\left|\frac{2h^L_i}{h_i} S^L_i\right|,|S^C_i|,\left|\frac{2h^R_i}{h_i} S^R_i\right|\right)
\end{equation}

\subsubsection{Local extrema}
\label{sec:lextrema}

Finally, the slope is set to $0$ in presence of a local extremum, only if it is considered as non smooth (see later):
\begin{equation}
    \tilde{S}_i =
    \begin{cases}
         0 \quad S^L_i \cdot S^R_i \leq 0\\
         \tilde{S}_i \quad \text{otherwise}. 
    \end{cases}
    \label{eq:extrema}
\end{equation}

\subsubsection{Predictor Corrector Scheme}
\label{sec:predictor}

In order to have a fallback scheme that is also second-order accurate in time, we use the predictor-corrector time integration strategy of the MUSCL-Hancock method \citep{van1984relation}. 
In what follows, for sake of simplicity, we describe our predictor step only for the one-dimensional advection equation:
\begin{equation}
    \partial_t u  + \partial_x (v_x u) = 0.
\end{equation}
The time evolution of cell $i$ is given in this case by:  
\begin{equation}
    \partial_t {u}^k_i = - (\bar{v}^k_{x,i}\partial_x {u}^k_i + {u}^k_i\partial_x {v}^k_{x,i}),
\end{equation}
where the superscript $k$ stands for the stage index of the ADER scheme. The spatial derivatives are obtained using any of our limited slope. As explained in \citet{van1984relation}, second order accuracy in time is achieved by advancing the cell interface values to the half time step $t^{k+\frac{1}{2}} = t^k + \frac{1}{2}w_k\Delta t$. 
The resulting predicted interface values are:
\begin{align}
    u^{R,k+\frac{1}{2}}_{i-1/2} = \bar{u}^k_{i} - \partial_x {u}^k_{i}\frac{\Delta x_i}{2} + \partial_t {u}^k_{i}\frac{w_k\Delta t}{2},\\
    u^{L,k+\frac{1}{2}}_{i+1/2} = \bar{u}^k_{i} + \partial_x {u}^k_{i}\frac{\Delta x_i}{2} + \partial_t {u}^k_{i}\frac{w_k\Delta t}{2}.
\end{align}
where the superscript $R$ means that this value is immediately to the right side of interface $i-1/2$ and $L$ immediately to the left side of interface $i+1/2$.
The Riemann problem is then solved at each interface $i+1/2$ making use of these predicted values:
\begin{align}
\label{eq:fb-flux}
    \hat{f}_{i+1/2} = RP\{{u}^{L,k+\frac{1}{2}}_{i+1/2},{u}^{R,k+\frac{1}{2}}_{i+1/2} \}.
\end{align}

\subsection{Troubled cell detection}

Now that we have described our fallback schemes, let us now describe the criteria we have used to identify a troubled cell and trigger the use of our fallback scheme.

\subsubsection{Numerical Admissibility Detection}

We start with our numerical criterion. As described in \citet{Vilar2019}, the numerical admissibility detection (NAD) criterion requires that the new control volume averaged solution satisfies a local maximum principle:
\begin{equation}
  \min(\bar{u}_{i-1}^{k},\bar{u}_{i}^{k},\bar{u}_{i+1}^{k}) \leq \bar{u}_{i}^{k+1} \leq \max(\bar{u}_{i-1}^{k},\bar{u}_{i}^{k},\bar{u}_{i+1}^{k}),  
  \label{eq:nad}
\end{equation}
that is, no new local extrema can be introduced. This criterion, as mentioned in \citet{Vilar2019}, is quite stringent, and does not distinguish between smooth and sharp extrema. We know that high-order methods excel at capturing smooth solutions, whereas low-order methods inherently dampen smooth extrema because of how slope limiters are designed. It is therefore necessary to apply the criterion \autoref{eq:nad} only away from smooth extrema.

\subsubsection{Smooth Extrema Detection}

Slope limiters must be avoided when a smooth extremum is present. In smooth extrema, the first derivative of the solution is a continuous function of space that crosses smoothly zero at the extremum position. Applying the \textit{moncen} criterion to the first derivative of the solution provides the means to discern between a smooth and a discontinuous extremum. 

We introduce a smoothness indicator $0 \leq \alpha \leq 1$, where $\alpha=1$ implies a smooth extrema. To compute this smoothness indicator we start by computing the discrete first order derivative of the control volume averaged solution using:
\begin{equation}
    {u}'_i = \frac{\bar{u}_{i+1}-\bar{u}_{i-1}}{\bar{x}_{i+1}-\bar{x}_{i-1}}.
\end{equation}
We then proceed with the \text{moncen} procedure for $u'$. The different slopes are now:
\begin{equation}
    S^C_i = \frac{u'_{i+1}-u'_{i-1}}{\bar{x}_{i+1}-\bar{x}_{i-1}}, \quad
    S^L_i = \frac{u'_{i}-u'_{i-1}}{\bar{x}_{i}-\bar{x}_{i-1}}, \quad
    S^R_i = \frac{u'_{i+1}-u'_{i}}{\bar{x}_{i+1}-\bar{x}_{i}}.
\end{equation}
The $\alpha$ coefficients at the left and right of cell $i$ are then defined as:
\begin{equation}
    \alpha^{L,R}_i = 
    \begin{cases}
        \min\left(1,\min\left(\frac{2h^{L,R}_i}{h_i} S^{L,R}_i,0\right)\frac{1}{S^C_i}\right) \quad &S^{C}_i < 0 \\
        1 \quad &S^{C}_i = 0\\
        \min\left(1,\max\left(\frac{2h^{L,R}_i}{h_i} S^{L,R}_i,0\right)\frac{1}{S^C_i}\right) \quad &S^{C}_i > 0. 
    \end{cases}
\end{equation}
Finally, we set the cell center value for $\alpha$ to be $\alpha_i = \min\left(\alpha^L_i,\alpha^R_i\right)$.

The new candidate high-order solution $\bar{u}_{i}^{k+1}$ is considered to be a smooth extremum if and only if:
\begin{equation}
   \min\left(\alpha^{k+1}_{i-1},\alpha^{k+1}_i,\alpha^{k+1}_{i+1}\right) = 1.
   \label{eq:smooth}
\end{equation}
This extra requirement extend the smooth extrema detection to the immediate neighboring subcells, so that a new local extremum according to  \autoref{eq:nad} can only be considered as a smooth extrema if the solution is smooth inside the local stencil.

In our extension to two-dimensional flows, an extremum is considered smooth if it satisfies the smoothness criterion  \autoref{eq:smooth} in both directions. Our 2D smooth extrema detection can be written as
\begin{align}
    &\alpha^{k+1}_x = \min\left(\alpha^{k+1}_{i-1,j},\alpha^{k+1}_{i,j},\alpha^{k+1}_{i+1,j}\right),\\
    &\alpha^{k+1}_y = \min\left(\alpha^{k+1}_{i,j-1},\alpha^{k+1}_{i,j},\alpha^{k+1}_{i,j+1}\right),\\
    &\min\left(\alpha^{k+1}_x,\alpha^{k+1}_y\right) = 1,
\end{align}
where $0 \leq j < N\times(p+1)$ is the subcell index along the second dimension of our spatial discretization. 

\subsubsection{Tolerance on the \textit{NAD} criterion}

There are cases where applying condition \autoref{eq:nad} might be counterproductive. For a uniform density, for example, tiny variations due to machine round-off errors can trigger the limiter. To circumvent this, we introduce a user defined tolerance parameter which allows small relative variations of each primitive variable as: 
\begin{equation}
  \bar{u}_{i,\min}^{k}-\varepsilon|\bar{u}_{i,\min}^{k}| \leq \bar{u}_{i}^{k+1} \leq \bar{u}_{i,\max}^{k}+\varepsilon|\bar{u}_{i,\max}^{k}|,  
\end{equation}
where $\bar{u}_{i,\min}^{k} = \min(\bar{u}_{i-1}^{k},\bar{u}_{i}^{k},\bar{u}_{i+1}^{k})$, $\bar{u}_{i,\max}^{k} = \max(\bar{u}_{i-1}^{k},\bar{u}_{i}^{k},\bar{u}_{i+1}^{k})$ and $\varepsilon$ is the tolerance factor. In this work, we use as fiducial value for the tolerance parameter $\varepsilon=10^{-5}$.

\subsubsection{Physical Admissibility Detection}

We also enforce as additional requirement that the new high order solution must be physically admissible, in other words we simply ask for the positivity of the gas density and the gas pressure. The physical admissibility detection (PAD) criterion is for  each control volume averaged density and pressure to satisfy:
\begin{align*}
    \bar{u}_{i} > u_{\rm min},
\end{align*}
where $u_{\rm min}$ is the minimum value allowed. Typically, we use here $u_{\rm min}=10^{-10}$. 

\subsection{Flux correction}

\begin{figure}
    \centering
    \includegraphics[width=0.8\columnwidth]{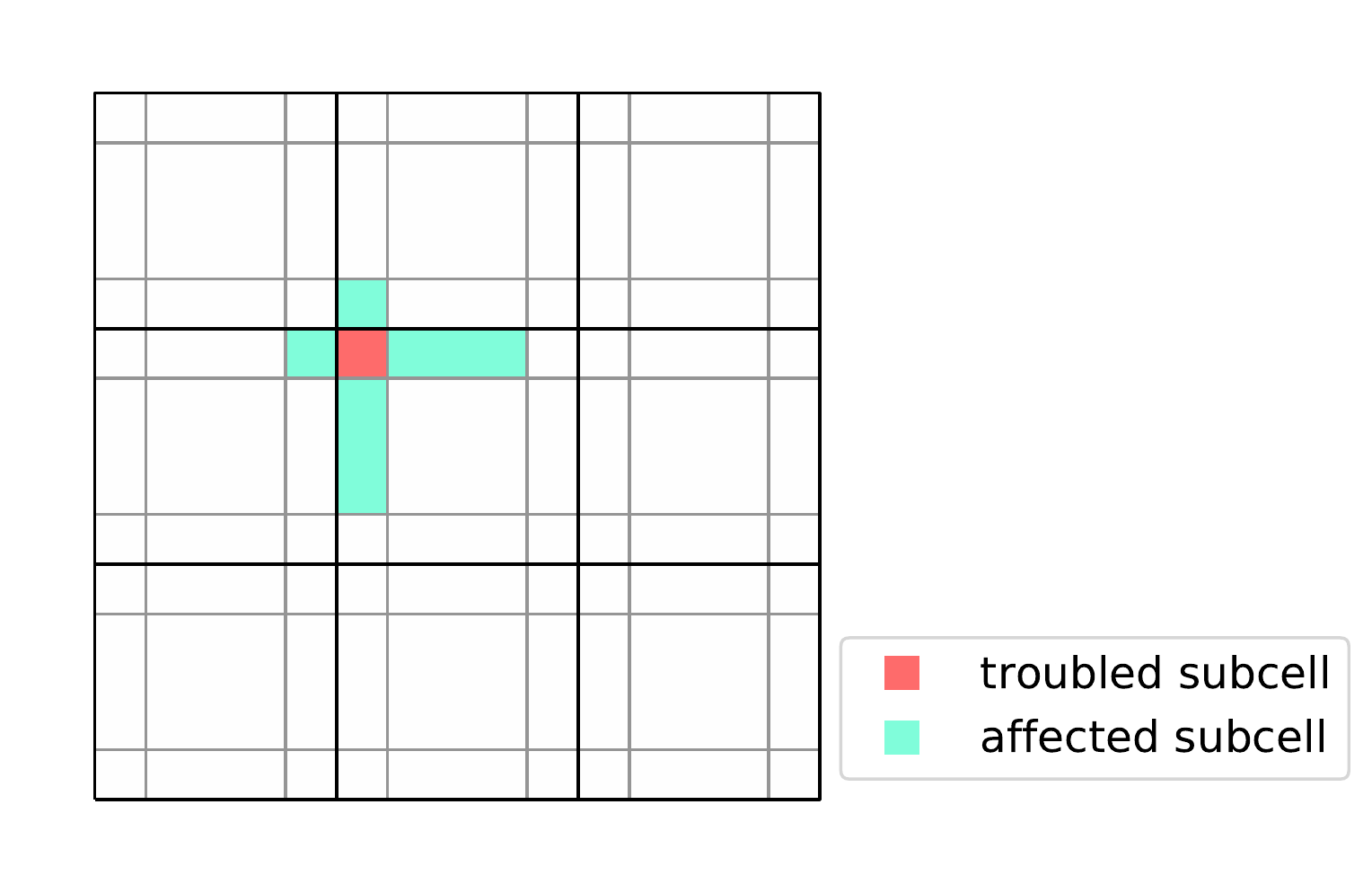}
    \caption{Visualization of the detection of a troubled subcell and the affected subcells in the flux correction.}
    \label{fig:3}
\end{figure}

In our a posteriori limiting strategy, any subcell that does not fulfill either the NAD or the PAD criteria, is considered as a troubled cell. The fluxes of this subcell are to be recomputed using our monotonicity and positivity preserving fallback scheme. As these fluxes are shared with neighboring control volumes, who will be in turn affected by these flux modifications, as depicted in \autoref{fig:3}.

In our implementation, we store all high-order fluxes for all stages of the ADER update. We first flag the troubled subcells. 2 (4) high-order fluxes in 1D (2D) associated with the troubled cells will be replaced by the fluxes of the fallback scheme. We need to recompute the control volume averaged variables of all the subcells affected by these new fluxes, one stage at the time in the ADER time quadrature as
\begin{equation}
    \bar{u}^{k+1}_{i} = \bar{u}^k_{i} - {\left(\hat{f}^{f,k}_{i+1/2}-\hat{f}^{f,k}_{i-1/2}\right)}\frac{w_k\Delta t}{h_{i}},
\end{equation}
where now $\hat{f}$ represent either a high-order flux or a fallback scheme flux. This strategy ensures that the overall scheme is strictly conservative.

\subsection{Summary of the Algorithm}
In \autoref{algo} we present a brief summary of the steps involved in the algorithm for the 1-dimensional advection equation scenario:
\begin{algorithm}
\SetAlgoLined
-Initialize data at control-volume averages $\bar{{u}}^0(\bar{x})$.\\
\While{$t<t_{end}$}{
-Transform $\bar{u}(\bar{x}) \to u(x^s)$.\\
-Compute $\Delta t$.\\
-Initialize predictions $\{u^{q,0}(x^s)\}^p_{q=0} = u(x^s)$.\\
\For{$q = 0, ... p$}{
    $\{u^{k,q}(x^s)\}^p_{k=0} \to \{u^{k,q}(x^f)\}^p_{k=0}$.\\
    $\{u^{k,q}(x^f)\}^p_{k=0} \to \{f^{k,q}(x^f)\}^p_{k=0}$.\\
    -Riemann solver at interfaces between elements $\{\hat{f}^{k,q}(x^f)\}^p_{k=0}$.\\
    \If{q<p}{
    -Update predictions
    \{$u^{k,q+1}(x^s)\}^p_{k=0}$.}
}
-Predicted fluxes $\{\hat{f}^{k,p}(x^f)\}^p_{k=0}$.\\
-Transform $u(x^s) \to \bar{u}(\bar{x})$.\\
\For{$k = 0, ... p$}{
-Compute candidate solution:\\
$\tilde{u}^{k+1}_i = \bar{u}^k_{i} - {\left(\hat{f}^{f,k}_{i+1/2}-\hat{f}^{f,k}_{i-1/2}\right)}\frac{w_k\Delta t}{h_{i}},$\\
\If{$\tilde{u}^{k+1}_{i}$ {\rm is~\emph{troubled}}}
{
    -Recompute $\hat{f}^{f,k}_{i\pm1/2}$ with fall-back scheme.
    -Flag $\tilde{u}^{k+1}_{i\pm1}$ as \emph{affected}.
}
-Compute corrected solution for \emph{troubled} and \emph{affected} control-volumes.\\
}
$t=t+\Delta t$.
}
\caption{SD method with \textit{a posteriori} limiting}
\label{algo}
\end{algorithm}

\section{Numerical results}
\label{sec:results}

In this section we present a series of numerical test to validate our implementation, and assess the performance of the method. We will also present, when needed, the additional components of the numerical scheme for 2D advection, 1D Euler equations and finally 2D Euler equations.

\subsection{Modifications of the algorithm for solving the 2D advection equation}

We have presented the algorithm so far for the 1D advection equation. We present in this section the necessary modifications to adapt the method to the 2D advection equation which writes in conservative form as
\begin{equation}
\partial_t \rho + \partial_x f( \rho ) + \partial_y g( \rho ) = 0, 
\end{equation}
where the fluxes are defined by
\begin{equation}
f(\rho)=\rho v_x~{\rm and}~g(\rho)=\rho v_y
\end{equation}
and the velocity field $\boldsymbol{v}=(v_x,v_y)$ is considered fixed in space and time. 
In the following we present the extension of the MUSCL-Hancock method (described in Sec.~\ref{sec:predictor}) for the two-dimensional case.
The predicted values interpolated at the left and right interfaces along the $x$-axis are given by:
\begin{align}
    u^{k+\frac{1}{2}}_{i-1/2,j} = \bar{u}^k_{i,j} - \partial_x u^k_{i,j}\frac{\Delta x_i}{2} + \partial_t u^k_{i,j}\frac{w_k\Delta t}{2},\\
    u^{k+\frac{1}{2}}_{i+1/2,j} = \bar{u}^k_{i,j} + \partial_x u^k_{i,j}\frac{\Delta x_i}{2} + \partial_t u^k_{i,j}\frac{w_k\Delta t}{2},
\end{align}
while the predicted values interpolated at the top and bottom interfaces along the $y$-axis are given by:
\begin{align}
    u^{k+\frac{1}{2}}_{i,j-1/2} = \bar{u}^k_{i,j} - \partial_y u^k_{i,j}\frac{\Delta y_i}{2} + \partial_t u^k_{i,j}\frac{w_k\Delta t}{2},\\
    u^{k+\frac{1}{2}}_{i,j+1/2} = \bar{u}^k_{i,j} + \partial_y u^k_{i,j}\frac{\Delta y_i}{2} + \partial_t u^k_{i,j}\frac{w_k\Delta t}{2}.
\end{align}
The predicted values $u^{k+1/2}_{i\pm1/2,j}$ and $u^{k+1/2}_{i,j\pm1/2}$ are then used to solve one-dimensional Riemann problems at the corresponding interfaces, providing respectively unique face average fluxes $\hat{f}^{k+1/2}_{i\pm1/2,j}$ and $\hat{g}^{k+1/2}_{i,j\pm1/2}$. 
Finally, the conservative update at time $t^{k+1}$ is written as:
\begin{equation}
    \bar{u}^{k+1}_{i,j} = \bar{u}^k_{i,j} 
    -\left(\frac{\hat{f}^{k+1/2}_{i+1/2,j}-\hat{f}^{k+1/2}_{i-1/2,j}}{\Delta x_{i}}\
    -\frac{\hat{g}^{k+1/2}_{i,j+1/2}-\hat{g}^{k+1/2}_{i,j-1/2}}{\Delta y_{i}}\right)w_k\Delta t.
\end{equation}

\subsubsection{2D advection of a square}

We first show the classical test of the advection of a square. The setup is as follows: 
\begin{equation*}
\rho = 
\begin{cases}
      2 \quad 0.25<x,y<0.75\\
      1 \quad \text{otherwise},
    \end{cases}
    v_x = 1, \quad v_y=1.
\end{equation*}
The computational volume is defined by $x,y \in [0,1]$ with periodic boundary conditions. 
For this test, any unlimited high-order method would exhibit oscillations triggered by the initial discontinuities. 
We show here that our a posteriori limiting strategy is capable of suppressing these oscillations and maintain both the monotonicity and the positivity of the solution. 

\begin{figure}
    \centering
    \includegraphics[width=\columnwidth]{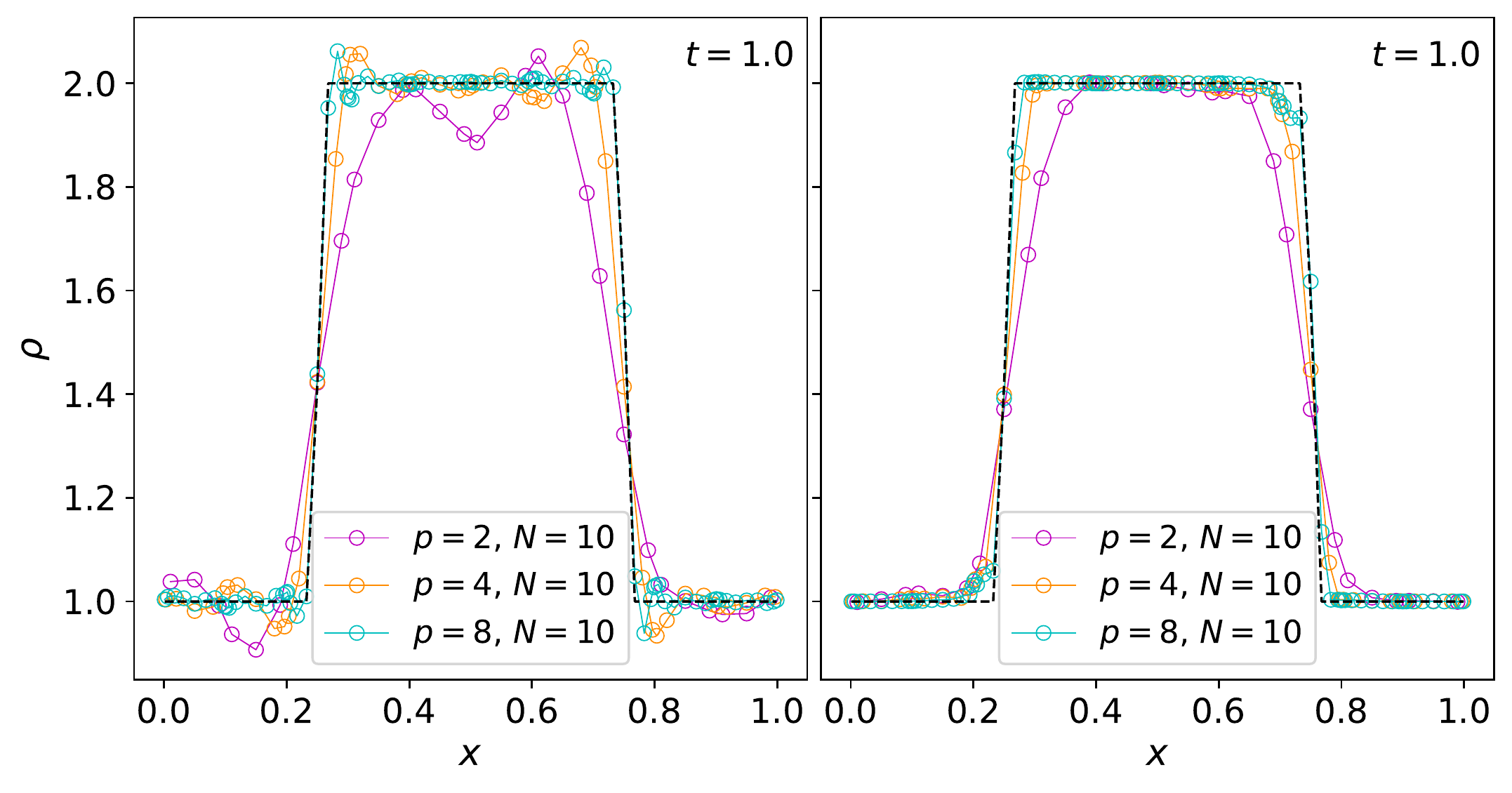}
    \caption{Control volume (subcell) averaged solution for the 2D advection of a square. In the left panel, we show the results without limiting, with the usual  oscillatory behavior of high-order methods. In the right panel, we show the results of SD with {\it a posteriori} limiting and the suppression of these oscillations. The black dashed line shows the exact solution at $t=1$.}
    \label{fig:square}
\end{figure}

In the left panel of \autoref{fig:square}, we present the results of the unlimited SD scheme, while in the right panel, we show the results of the SD method with a posteriori limiting, both at $t=1$ and for polynomial orders $p=2,4$ and $8$, all of them with only $10$ elements per dimension. The results without limiting exhibit wild oscillations in all three cases, in contrast to the limited method with no oscillations in all three cases. The density is nicely bounded by the initial range $\rho \in [1,2]$, which support the fact that our scheme satisfies a maximum principle and is therefore positivity preserving. 

We observe that even for such a small number of elements, the initial top hat profile has been barely affected by numerical diffusion. The capability of our method to preserve the sharpness of the initial discontinuities inside the elements is quite impressive. This sub-cell limiting property explains why the quality of the results is still increasing with increasing order of accuracy, even in presence of a discontinuity. 

Our results for the advection of a square compare very well with the ones obtained by \citet{Vilar2019}. They used a classical DG scheme with a posteriori limiting using like us $p+1$ control volumes for their fallback scheme. They however used a flux projection method to project the FV flux to the DG weak formulation. This distinction between SD and DG turns out to be a minor difference for this particular test. 

As shown by \citet{Vilar2019}, classical DG schemes using mode by mode limiters such as the one proposed by \citet{krivodonova2007limiters} perform much worse for very high order polynomial degree in presence of sharp discontinuities than our current FV fallback strategy.

\subsubsection{2D advection of a sine wave}

The second test is the 2D advection of a smooth sine wave. 
The setup is as follows:
\begin{equation}
\label{eq:sine_ic}
\rho = \sin(2\pi(x + y)), \quad v_x = 1, \quad v_y=1.
\end{equation}
The computational volume is defined by $x,y \in [0,1]$ with periodic boundary conditions. 
The simulation of a smooth solution is a problem at which high-order methods traditionally excel. 
We want to test here that our a posteriori limiting technique is not detrimental to the quality of the solution. 

\begin{figure}
    \centering
    \includegraphics[width=\columnwidth]{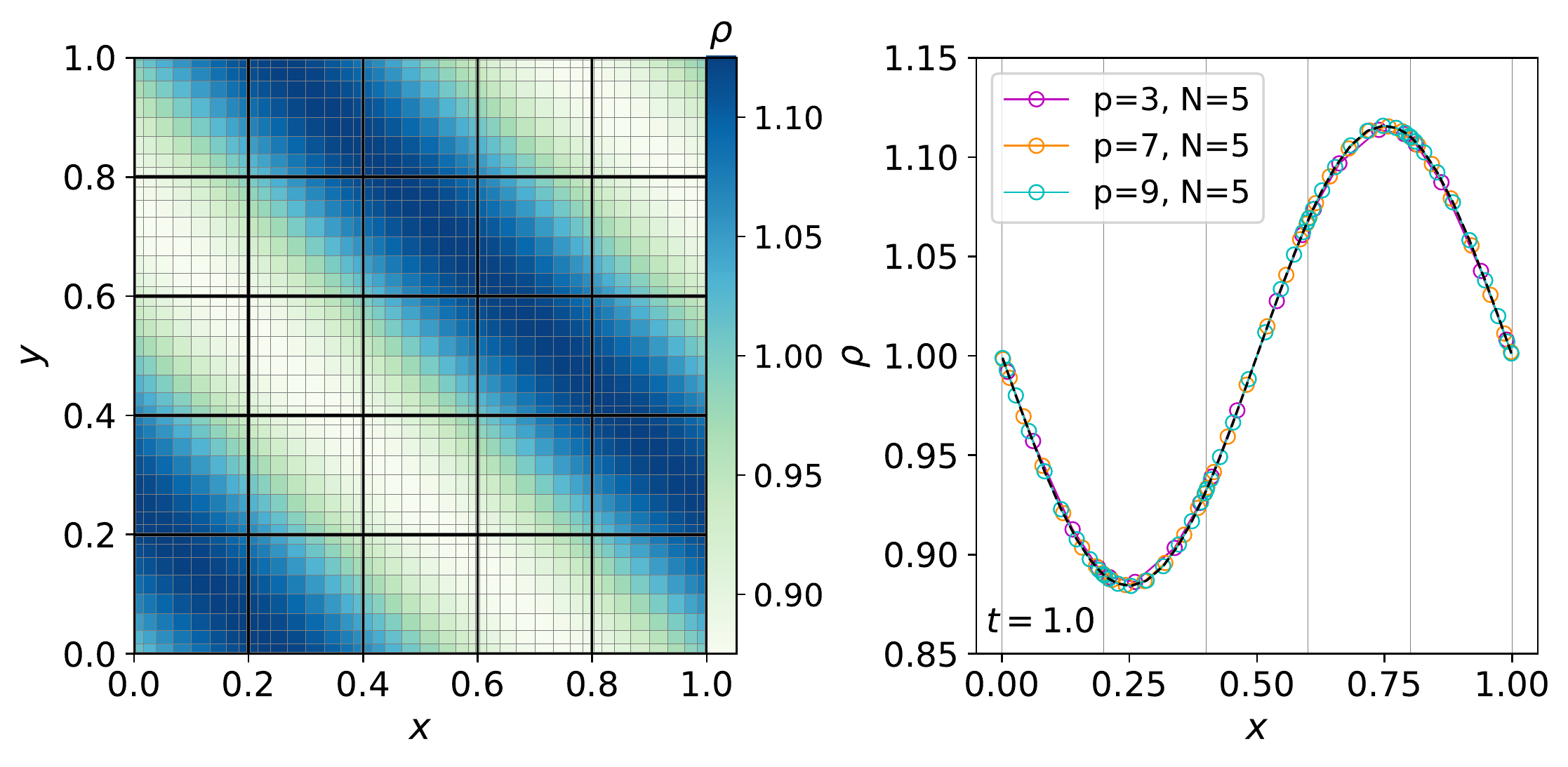}
    \caption{Control volume (subcell) averaged solution for the 2D advection of a sine wave at $t=1$ for a mesh with $5\times5$ elements. On the left panel, we show a color map of the solution for polynomial degree $p=9$. On the right panel, the results for $p=3, 7$ and $9$ are interpolated at $y=0.5$. The black dash line shows the exact solution at $t=1$. }
    \label{fig:sine}
\end{figure}

In \autoref{fig:sine}, the left panel shows a snapshot of the density at $t=1$, corresponding to one complete translation in both directions, for $p=7$ and $N=5$. 
Both element and control volume boundaries are shown respectively as black and grey lines. 
On the right panel, we present a cut along the $x$-axis at $y=0.5$ and $t=1$, for $n=3,7$ and $9$ and only $5$ elements. 
We want to stress that the smooth solution is preserved for all three cases, validating our smooth extrema detection algorithm.

\begin{figure}
    \centering
    \includegraphics[width=\columnwidth]{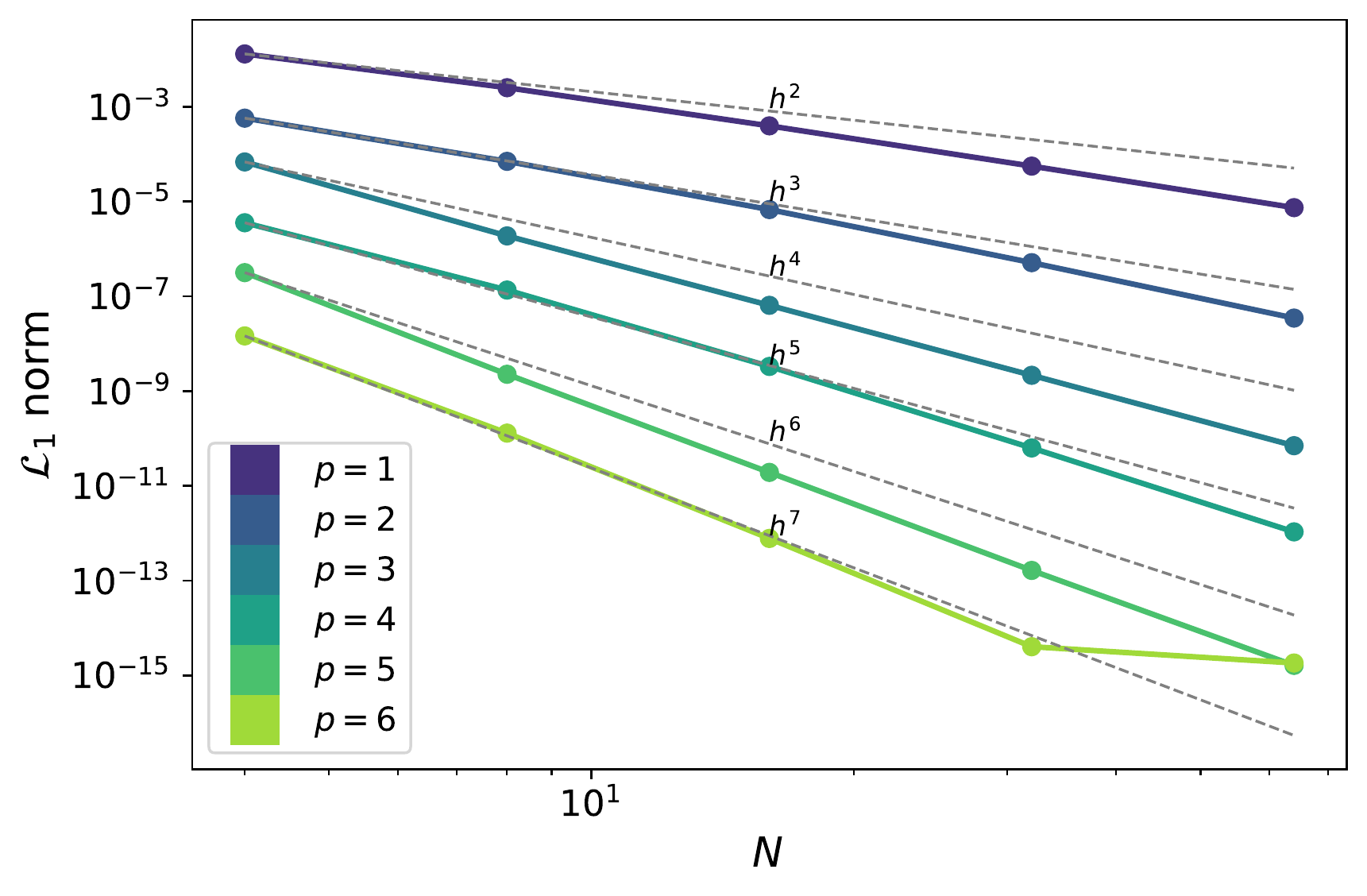}
    \caption{$L_1$ norm of the error for the 2D advection of the sine wave at different polynomial degrees and element resolution.    \label{fig:convergence_rates}
    }
\end{figure}

In \autoref{fig:convergence_rates}, the measured and expected L1 error is shown as a function of element resolution for various polynomial degrees, from $p=1$ to $p=7$. We note that for $p=6$ and $N=64$ in each dimension, our error reaches the level of machine precision. 

\subsubsection{Rotation of a slotted disc}

We study now the performance of the method for the rotation of a slotted disc. This classical test features a non-uniform velocity field together with a discontinuous solution. The setup is as follows:
\begin{align*} 
\rho &=
\begin{cases}
      2 \quad 0.25<x_c^2+y_c^2<0.15^2, \quad
        |x_c| > 0.025, \quad y > 0.85  \\
      1 \quad \text{otherwise},
    \end{cases}\\
v_x &= -y_c, \quad v_y=x_c,
\end{align*}
where $x_c= x-0.5$ and $y_c= y-0.5$, in a computational volume defined by $x,y \in [0,1]$ with zero gradient boundary conditions.

\begin{figure}
    \centering
    \includegraphics[width=\columnwidth]{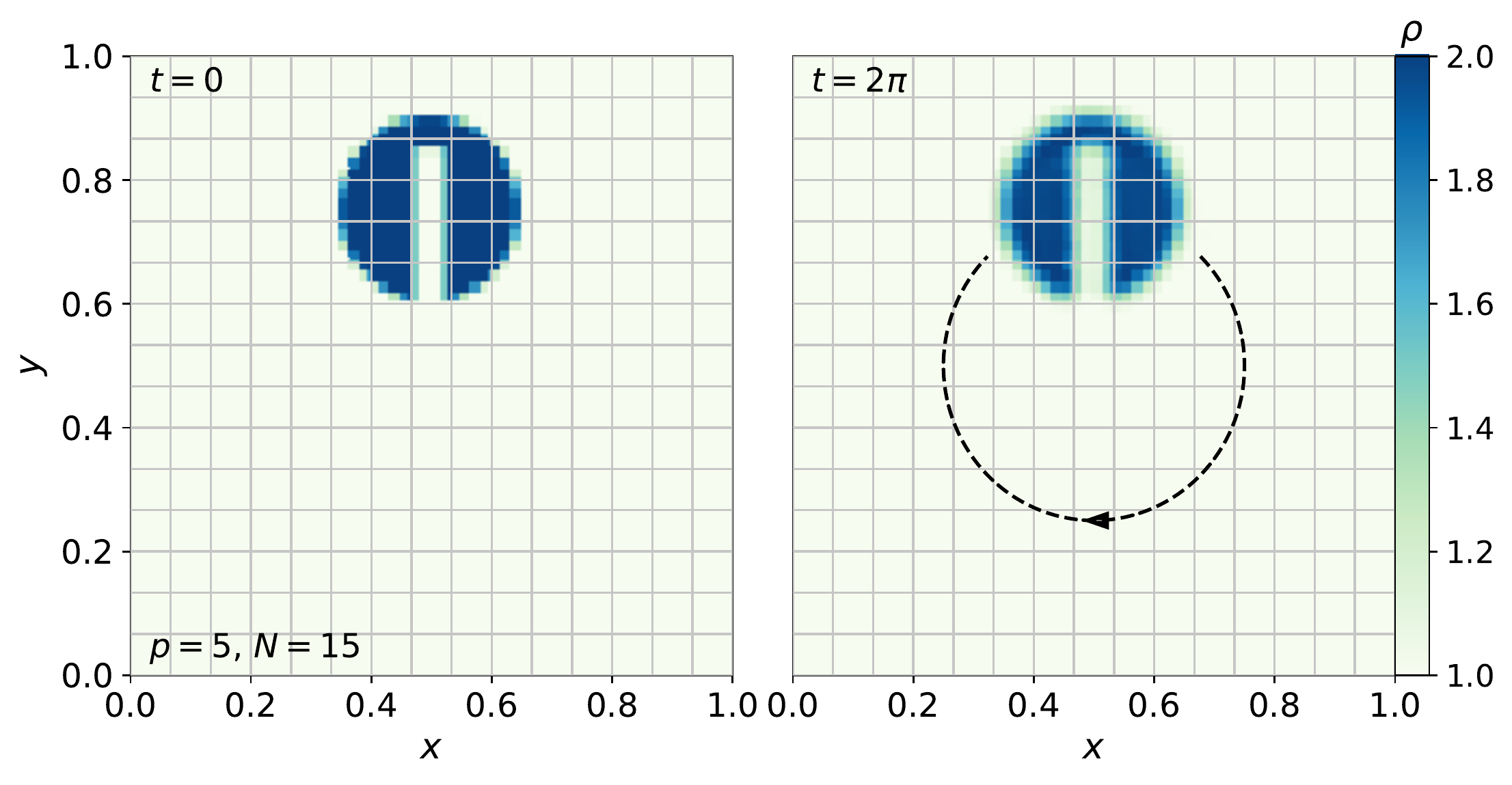}
    \caption{Rotation of a slotted disc. We present the control volume (subcell) averaged solution for a simulation with a polynomial degree of $p=5$ and a mesh with $15\times15$ elements.  The left panels shows the initial condition, and the right panel shows the numerical solution after one complete revolution.}
    \label{fig:pacman}
\end{figure}

In \autoref{fig:pacman} we present color maps for the control volume averaged solution of a simulation with $p=5$  and  $15\times15$ elements.  The left panel shows the initial condition, while the right panel shows the numerical solution after one complete rotation ($t=2\pi$). With such a small number of elements, the slotted disc is captured almost exclusively by the internal degrees of freedom (or subcells) within the elements.  The snapshot at $t=2\pi$ shows that the method is able of capturing this discontinuous profile with very little numerical diffusion and with details at scales smaller than the size of the elements. Our results are very similar to the one found by the DG scheme of \citet{Vilar2019} using the same polynomial degree $p=5$ and the same number of elements $N=15$,

\subsection{Modifications of the algorithm for the 1D Euler equations}

We now present our method when applied to the solution of the 1D Euler equations. The Euler equations are the prototype of nonlinear hyperbolic systems of conservation laws, governing the motion of inviscid and compressible ideal fluids. We write the Euler equation in vector form for the one-dimensional case as:
\begin{equation}
    \partial_t \mathbf{U} + \partial_x \mathbf{F}(\mathbf{U}) = 0 
\end{equation}
where the solution vector $\mathbf{U}$ and the flux vector $\mathbf{F}$ are respectively
\begin{equation}
\mathbf{U} = 
    \begin{pmatrix}
    \rho \\
    \rho v_x \\
    E
    \end{pmatrix}, \quad
\mathbf{F} = 
    \begin{pmatrix}
    \rho v_x \\
    \rho v^2_x + P \\
    (E + P)v_x
    \end{pmatrix},    
\end{equation}
where $\rho$ is the mass density, $v_x$ the x-velocity, $E = e + \frac{1}{2}\rho v^2_x$ the total energy, equal to the sum of the internal energy density $e$ and the kinetic energy density. Finally, the system is closed with the equation of state of an ideal gas $P = (\gamma-1)e$, with $\gamma$ the constant adiabatic index.

For our present SD implementation,  the interpolation of state variables to the flux points is performed using the conservative variables $\mathbf{U}$. For the fall-back scheme, on the other hand, the interpolation to control volume faces is performed using the primitive variables $\mathbf{W}$ and the following quasi-linear form:
\begin{equation}
\mathbf{W} = 
    \begin{pmatrix}
    \rho \\
    v_x \\
    P
    \end{pmatrix}, \quad
\mathbf{\partial_t W} = -
    \begin{pmatrix}
    v_x\partial_x \rho + \rho\partial_x v_x\\
    v_x\partial_x v_x +\frac{1}{\rho}\partial_x P\\
    v_x\partial_x P +\gamma P \partial_x v_x
    \end{pmatrix}.
\end{equation}
We compute face average values of the primitive variables at the half time step $t^{k+\frac{1}{2}}$  as:
\begin{align}
    \mathbf{W}^{k+\frac{1}{2}}_{i-1/2} &= \overline{\mathbf{W}}_{i} - \partial_x\mathbf{W}_{i}\frac{\Delta x_i}{2} + \partial_t \mathbf{W}_{i}\frac{w_k\Delta t}{2},\\
    \mathbf{W}^{k+\frac{1}{2}}_{i+1/2} &= \overline{\mathbf{W}}_{i} + \partial_x\mathbf{W}_{i}\frac{\Delta x_i}{2} + \partial_t \mathbf{W}_{i}\frac{w_k\Delta t}{2}.
\end{align}
where $\overline{\bf W}$ represents the control volume averaged primitive variables  computed using the control volume averaged conservative variables $\overline{\bf U}$. These predicted face values for primitive variables are used to solve the Riemann problem at each control-volume interface and obtain the fluxes to update the conservative variables. In our current implementation, we use the Local Lax Friedrich (LLF) Riemann solver for the SD scheme, and the HLLC Riemann solver for the fallback scheme \citep{toro2013riemann}.

\subsubsection{Sod shock tube test}

\begin{figure}
    \centering
    \includegraphics[width=\columnwidth]{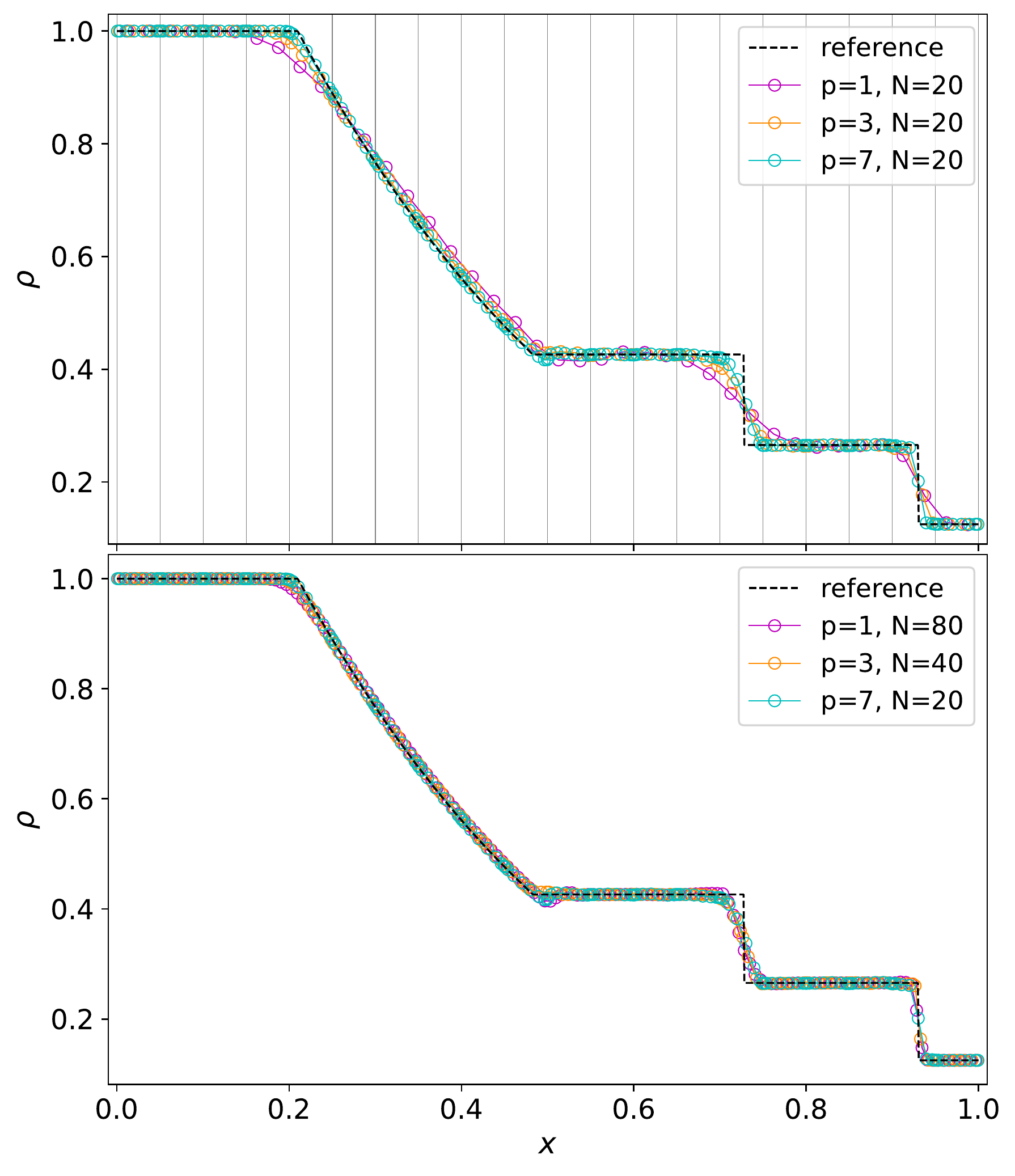}
    \caption{Control volume (subcell) averaged solution for the Sod shock tube test \citep{sod1978survey} at $t=0.245$. The  top panel shows the results for $N=20$ and increasing polynomial degree $p=1,3$ and $7$.  The bottom panel shows the results for $N=80$, $40$ and $20$ and $p=1,3$ and $7$, keeping the total number of DOF constant and equal to   $N(p+1)=160$. The black dashed line shows the reference solution computed with $1024$ cells and a second order Godunov method.}
    \label{fig:sod}
\end{figure}

This hydrodynamical test presents all 3 nonlinear waves for ideal fluids, namely a rarefaction wave, a contact discontinuity, and a shock wave as described for example in \citet{sod1978survey}. The initial conditions of this classical problem are as follows:
\begin{align*} 
\rho &= 
\begin{cases}
     \quad 1.0 \quad x<0.5 \\
     0.125 \quad x>0.5,
    \end{cases}
P = 
\begin{cases}
     1.0 \quad x<0.5 \\
     0.1 \quad x>0.5,
    \end{cases} v_x = 0,
\end{align*}
in a box of size $L=1$ with zero gradient boundary conditions. In \autoref{fig:sod}, we present the control volume averaged solution at time $t=0.245$. The top panel of \autoref{fig:sod} shows the results with only $20$ elements and increasing polynomial degree ($p=2$, $4$ and $8$). We show the boundaries of the elements as the thin vertical grey lines. 

We clearly see the improvement in the quality of the solution as the order is increased. This is particularly true for the rarefaction wave, where high order methods usually excel. This is also true for the contact discontinuity. This less trivial result is a consequence of our subcell a posteriori limiting strategy, allowing to exploit the internal DOF of the elements, even in presence of a discontinuity. 

The bottom panel of \autoref{fig:sod} shows our results for different simulations that maintain constant the total number of DOF defined here as $N(p+1)=160$. We therefore use a decreasing resolution $N=80$, $40$ and $20$ but with increasing polynomial degree with $p=1$, $3$ and $7$. The contact discontinuity benefits from the increased order of accuracy, but more importantly, the discontinuities are not suffering from the decreased resolution, which is usually a weak point of traditional FE methods \citep[see discussion in][]{Vilar2019}.

We note that our results for the Sod test are very similar to the one reported in \citet{Vilar2019}. In particular, our results with $p=4$ and $N=20$ match closely the one in  \citet{Vilar2019} for $p=8$ and $N=10$. This seems to indicate that two different FE methods (DG and SD) based on a similar a posteriori limiting approach, compare very well for a simulation featuring the 3 typical waves in compressible fluid dynamics. 

In order to compare to FV high-order methods, we refer the reader to the work of \citet{zhao2017new}, in which the results of the Sod test compare very well with our results. This seems to indicate that a FE method such as SD with a posteriori limiting is competitive with high-order WENO schemes. This is not necessarily always true as will be discussed later.

\subsubsection{Woodward and Colella interacting blast waves test}

\begin{figure}
    \centering
    \includegraphics[width=\columnwidth]{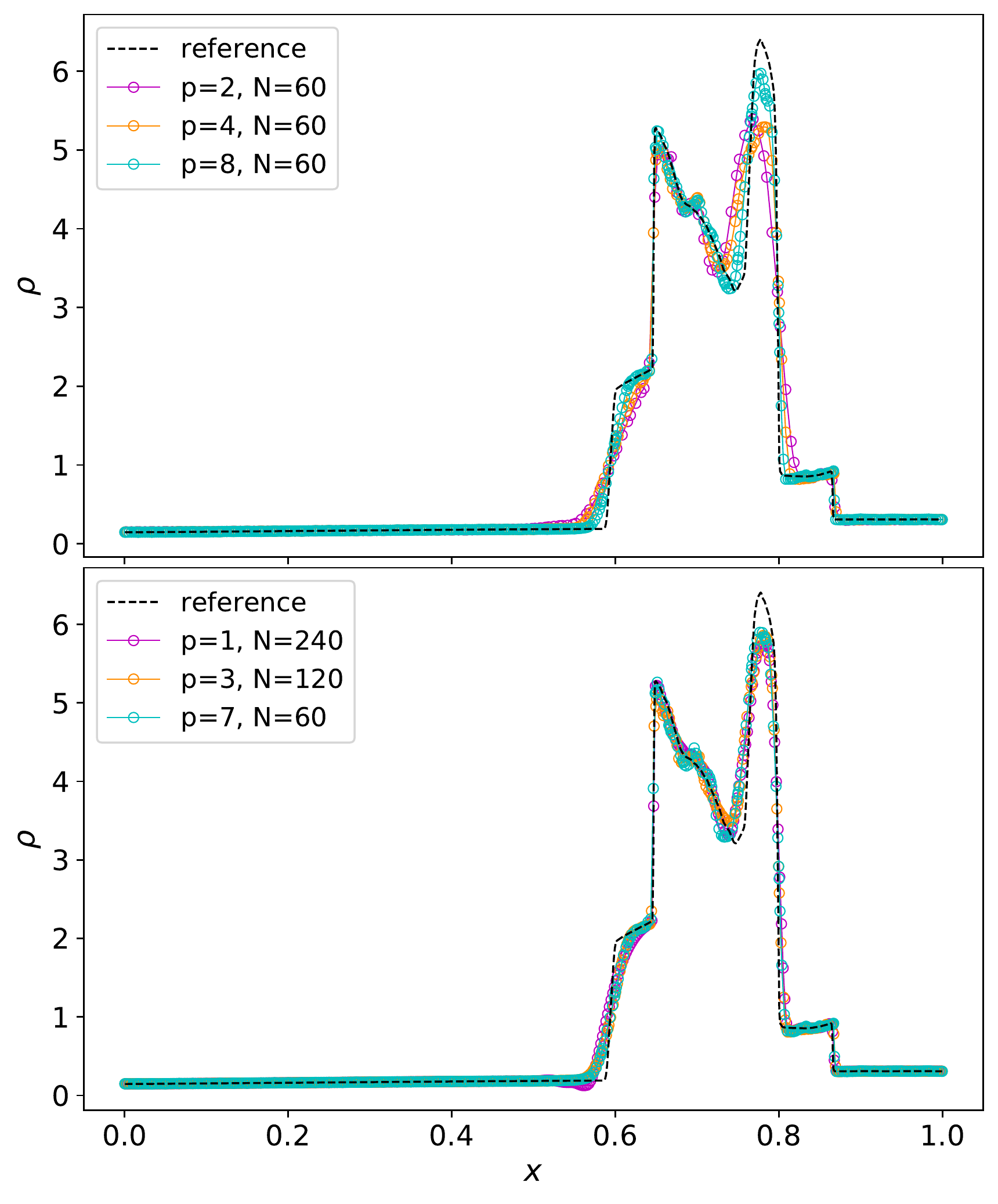}
    \caption{Control volume (subcell) averaged density for the blast wave test of \citet{woodward1984numerical} at $t=0.038$. The top panel shows the results for $N=60$ and increasing polynomial degree $p=2,4$ and $8$. The bottom panel shows the results for $N(p+1)=480$ for $p=1,3$ and $7$. The black dashed line show a reference solution obtained with $1000$ cells and a second-order Godunov method.}
    \label{fig:blast}
\end{figure}

This other classical test described in \citet{woodward1984numerical} features two strong shocks propagating towards the center of a box of size $L=1$, with two strong rarefaction waves propagating in the opposite direction, towards the left and right boundaries of the domain with wall (reflective) boundary conditions. The initial conditions of this test are as follows:
\begin{align*} 
\rho &= 1,\quad
P = 
\begin{cases}
     10^3 \quad x\leq0.1 \\
     10^{-2} \quad 0.1 < x \leq 0.9 \\
     10^2 \quad x>0.9,
    \end{cases} v_x = 0,
\end{align*}
In \autoref{fig:blast} we present the results of our various simulations at $t=0.038$. The top panels shows the control volume averaged solution for increasing polynomial degree $p = 2, 4$ and $8$ and for a fixed resolution $N=60$. The bottom panel shows the results of another series of simulations with increasing polynomial degree $p = 1, 3$ and $7$ but keeping the number of DOF constant and equal to $N(p+1)=480$. 

It is clear that our a posteriori limiting strategy allows us to resolve the sharp discontinuities without triggering spurious oscillations. We see that for a fixed number of elements, the quality of the solution is increased with increasing order of accuracy. Since we are dealing with discontinuities, this increased quality is mostly due to the larger number of DOF we can exploit via our FV fallback scheme. 

Using the DG method with mode by mode limiting, \cite{krivodonova2007limiters} obtained very similar results with $N=200$ and $p=2$ or $N=400$ and $p=1$, in agreement with our results with similar numbers of DOF. Note however that our simulation with $N=60$ and $p=7$ is still improving the solution, while very high order solutions are not presented in  \cite{krivodonova2007limiters}. This again supports the idea of \citet{Vilar2019} that mode by mode limiting is not competitive at very high order and polynomial degree, compared to a posteriori limiting with a FV fallback scheme.

The top panel of \autoref{fig:blast}  can be compared exactly to Fig.~26 of \citet{Vilar2019} with exactly the same parameters 
$N=60$ and $p=2$, $4$ and $8$. Our SD results seem slightly better than the one presented there using DG. Note that \citet{Vilar2019} used a RK3 time integrator, while in our case we use ADER with the same accuracy in time than in space. This could explain this apparent discrepancy in this particularly difficult case. 

\citet{zhao2017new} however obtained significantly better results with their 8th order WENO scheme using 400 mesh points than our simulation with $p=7$ and $N=60$. Although our number of DOF is higher with $N(p+1)=480$, our results are not as good. One possible explanation is that our underlying mesh is non-uniform, with SD flux points being squeezed towards the left and right boundaries of the elements. This results into a lower effective resolution in our case than for a FV method with a uniform mesh.

\subsubsection{Shu and Osher shock tube test}

\begin{figure*}
    \centering
    \includegraphics[width=0.95\textwidth]{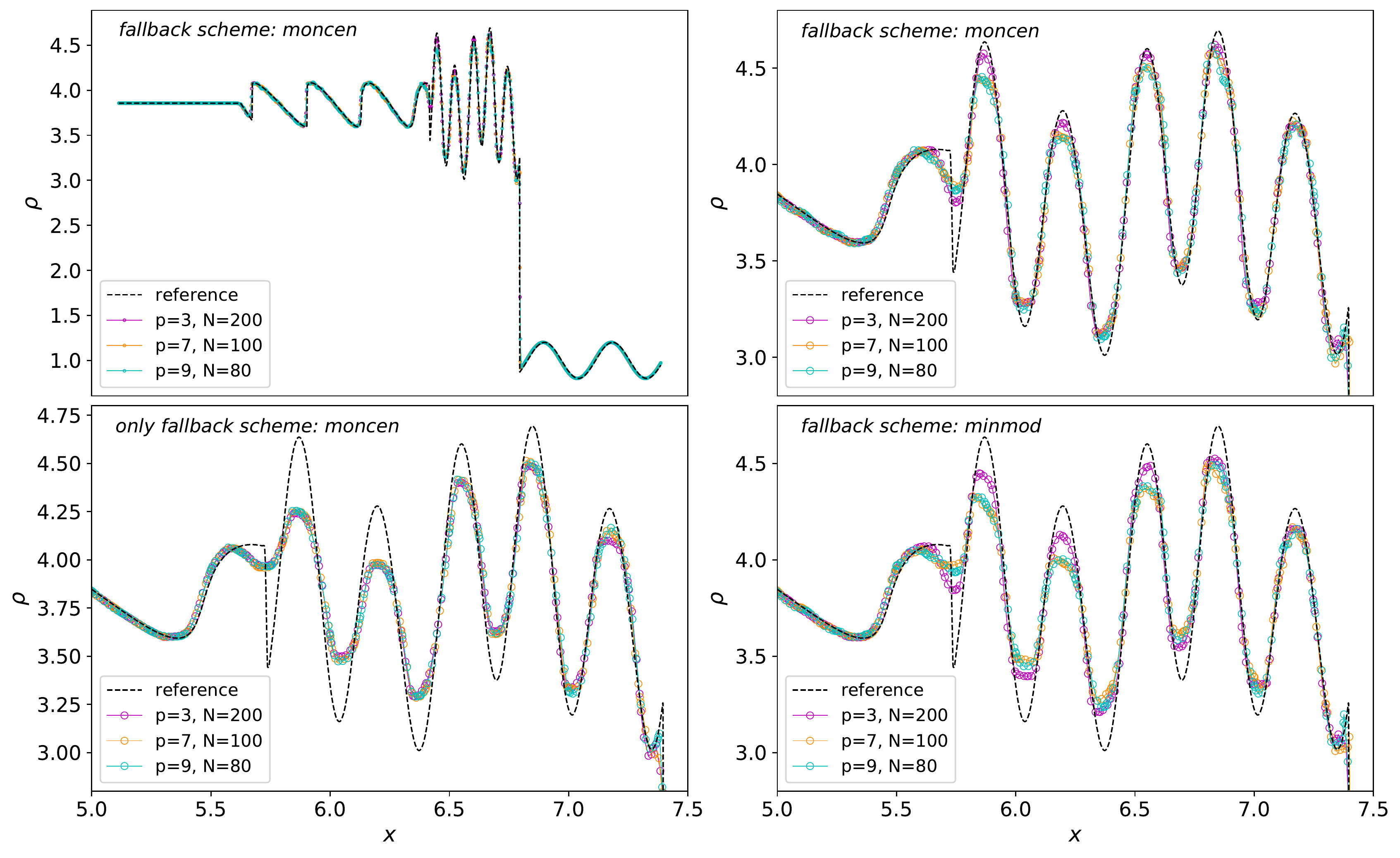}
    \caption{Density control volume averages of each subcell for the Shu-Osher test \citep{shu1988efficient} at $t=1.8$. On the top row the results with \textit{moncen} for the fallback scheme. The simulations where performed with a constant number of degrees of freedom ($N(p+1)=800$) for three values of the polynomial degree ($p=3,7 $ and $9$). On the left are shown the results for the whole domain, on the right a zoom in at $x\in[5,7.5]$.
    On the bottom right the results with \textit{minmod} for the fallback scheme are shown with same number of degrees of freedom. On the bottom left the results with only the \textit{moncen} fallback scheme are shown for $N=200$ and increasing order. The black dashed line is a high-resolution reference solution computed with $16384$ cells using our second order MUSCL scheme.}
    \label{fig:shu}
\end{figure*}

The Shu and Osher shock tube test \citep{shu1988efficient} features the complex interaction of a shock with the smooth profile of an acoustic wave. The initial conditions are as follows: 
\begin{align*} 
(\rho,P,v_x) = 
\begin{cases}
    (3.857143, 10.33333, 2.629369) \quad &x_c<-4 \\
    (1+0.2\sin(5x),1,0)  \quad &x_c>-4,
\end{cases}
\end{align*}
where $x_c= x-5$. The computational domain has dimension $L=10$ with zero gradient boundary conditions. 
This test is particularly difficult as the shock compresses the sinusoidal density profile. It remains smooth but with a very high spatial frequency that challenges the different components of the numerical schemes.

In \autoref{fig:shu}, we present our results at $t=1.8$ for different simulations with a fixed number of DOF $N(p+1)=800$, for polynomial degrees $p=3, 7$ and $9$. The top two panels show the results using the \textit{moncen} slope limiter for the fallback scheme, whereas the bottom right panel shows the results using the \textit{minmod} slope limiter for the fallback scheme. The bottom left panel shows the solution we obtain using only the fallback scheme with the moncen slope limiter. 

From a detailed comparison of these different results, we can conclude that the quality of the fallback scheme is of primary importance in our a posteriori limiting strategy. We also conclude that the added value of the high-order SD scheme is significant compared to using only the second-order fallback scheme. Finally, our results using the moncen fallback scheme and 800 DOF seems quasi-converged, with very similar results for the different polynomial orders. This supports our earlier conclusions for solution featuring discontinuities: if we go to high order while reducing the number of elements, our solution does not deteriorate. 

Our solution compares quite well with the one presented in \cite{krivodonova2007limiters}. This author used DG with mode by mode limiting with $N=250$ and $p=2$. An earlier and similar study by \citet{burbeau2001} found also similar results with DG and mode by mode limiting with $N=300$ and $p=3$. Our results are particularly close to the one presented in \citet{Vilar2019}, who used $N=50$ and $p=6$ for DG with a posteriori limiting. It emerges from these different studies that DG with mode by mode limiting is only competitive for moderate polynomial degrees (say $p\le3$), while a posteriori limiting appears more robust for very high order calculations with $p>3$. 

We have found however in the literature examples of schemes that outperform significantly our new method. First, \citet{Dumbser2014} report on a similar test with $N=40$ and $p=9$ (only 400 DOF) with a much higher quality than our simulation with $N=80$ and $p=9$ (800 DOF). In this study, the authors also use a posteriori limiting, but the fallback scheme is a third order WENO scheme with a Cartesian grid of $2p+1$ cells covering each troubled element. The resolution of their fallback scheme is therefore a factor of 2 better than here. In addition, they used a uniform mesh, to be compared to our non-uniform distribution of flux points. All these elements combined could explain why \citet{Dumbser2014} outperforms us.

A second striking example can be found in \citet{zhao2017new}. These authors used the same test to evaluate the performance of a novel class of WENO schemes while varying the degree of their reconstruction polynomials ($p=3$, $4$, $5$ and $7$). They also used only $400$ DOF (in their case mesh points). Our simulations shown in \autoref{fig:shu} all compare well with the fourth order ($p=3$) WENO scheme of \citet{zhao2017new}. On the other hand, the quality of the solutions with $p=4$, $5$ and $7$ in \citet{zhao2017new} increases steadily and significantly, which is not the case for us, as we keep the number of DOF constant. Another important difference (besides the one between FV and FE methods) resides in the limiting strategy, a priori limiting for \citet{zhao2017new} and a posteriori limiting for us. One interesting idea to improve the quality of our solution could be to use a hierarchy of fallback schemes of progressively lower and lower order, as explained in \citet{loubere2014new}. In our case, we directly fallback to second-order, which might cause the solution to degrade too rapidly in this test.

\subsection{Modifications of the algorithm for the 2D Euler equations}

We now present our method for the solution of the 2D Euler equations, written in vector form as:
\begin{equation}
    \partial_t \mathbf{U} + \partial_x \mathbf{F}(\mathbf{U}) + \partial_y \mathbf{G}(\mathbf{U}) = 0 
\end{equation}
where the solution vector $\mathbf{U}$ and the flux vectors $\mathbf{F}$ and $\mathbf{G}$ are now:
\begin{equation}
\mathbf{U} = 
    \begin{pmatrix}
    \rho \\
    \rho v_x \\
    \rho v_y \\
    E
    \end{pmatrix}, \quad
\mathbf{F} = 
    \begin{pmatrix}
    \rho v_x \\
    \rho  v^2_x + P \\
    \rho v_x v_y \\
    (E + P)v_x
    \end{pmatrix},  \quad
\mathbf{G} = 
    \begin{pmatrix}
    \rho v_y \\
    \rho v_x v_y \\
    \rho  v^2_y + P \\
    (E + P)v_y
    \end{pmatrix}. 
\end{equation}
Similarly to the 1D case, the interpolation of the state variables for the SD scheme, from the solution points to the flux points, is performed using the conservative variables, while the interpolation to the face averaged values for the fall-back scheme is performed using the primitive variables $\mathbf{W}$. In the next few paragraphs, we present a brief description of the MUSCL-Hancock method, used here as our fallback scheme, in a two-dimensional framework. The primitive variables and the corresponding quasi-linear form write:
\begin{align}
&\mathbf{W} = 
    \begin{pmatrix}
    \rho \\
    v_x \\
    v_y \\
    P
    \end{pmatrix}, \quad \\
&\mathbf{\partial_t W} = -
    \begin{pmatrix}
    (v_x\partial_x \rho + \rho\partial_xv_x) + (v_y\partial_y \rho +  \rho\partial_y v_y)\\
    (v_x\partial_x v_x +\frac{1}{\rho}\partial_x P) + v_y\partial_y v_x\\
    v_x\partial_x v_y + (v_y\partial_y v_y + \frac{1}{\rho}\partial_y P) \\
    (v_x\partial_x P +\gamma P\partial_x v_x) + (v_y\partial_y P + \gamma P\partial_y v_y)
    \end{pmatrix},  \quad
\end{align}
The face average values for the primitive variables in the x-direction are given by:
\begin{align}
    \mathbf{W}^{k+\frac{1}{2}}_{i-1/2,j} = \overline{\mathbf{W}}_{i,j} - \partial_x\mathbf{W}_{i,j}\frac{\Delta x_i}{2} + \partial_t \mathbf{W}_{i,j}\frac{w_k\Delta t}{2},\\
    \mathbf{W}^{k+\frac{1}{2}}_{i+1/2,j} = \overline{\mathbf{W}}_{i,j} + \partial_x\mathbf{W}_{i,j}\frac{\Delta x_i}{2} + \partial_t \mathbf{W}_{i,j}\frac{w_k\Delta t}{2}.
\end{align}
whereas for the y-direction we have:
\begin{align}
    \mathbf{W}^{k+\frac{1}{2}}_{i,j-1/2} = \overline{\mathbf{W}}_{i,j} - \partial_y\mathbf{W}_{i,j}\frac{\Delta y_i}{2} + \partial_t \mathbf{W}_{i,j}\frac{w_k\Delta t}{2},\\
    \mathbf{W}^{k+\frac{1}{2}}_{i,j+1/2} = \overline{\mathbf{W}}_{i,j} + \partial_y\mathbf{W}_{i,j}\frac{\Delta y_i}{2} + \partial_t \mathbf{W}_{i,j}\frac{w_k\Delta t}{2}.
\end{align}
These predicted states are then used to solve for a 1D Riemann problem at each control volume interfaces. The resulting fluxes are use to update the conservative variables. As for the 1D case, we use the LLF Riemann solver for the SD update
and the HLLC Riemann solver for the fallback scheme \citep{toro2013riemann}.

\subsubsection{Nonlinear sound wave test}

A classical smooth solution of the Euler equations are adiabatic sound waves. They
are however exact solutions of the {\it linearized} Euler equations. In the general case, nonlinear effects such as higher harmonic generation are expected, unless one uses very small amplitude waves. 

In the course of this work, we realized
that high order methods are particularly challenging to use in this context because errors are so small. Nonlinear corrections must be included in the exact reference solution if one wants to compute accurately the $\mathcal{L}_1$ norm of the error of the numerical solution. 

This is why we have carefully designed a new test using the perturbative second-order solution of the propagation of a planar wave, leading to a small but detectable secondary wave with twice the frequency of the initial planar wave.
Our setup consist of a nonlinear acoustic perturbation over a uniform equilibrium state:
\begin{align*}
    \rho &= \rho_0 + \rho', \quad v = v',\quad P = P_0 + P',
\end{align*}
with the initial perturbation at $t=0$ given by:
\begin{align*}
    v'(x) &= A\sin(kx), \quad
    \rho'(x) = \rho_0 v'/c_{s,0}, \quad
    P'(x) = \gamma P_0 v' /c_{s,0},
\end{align*}
where $k= 2\pi/\lambda$. We adopt here $\lambda = L$. 
The corresponding time-dependent solution for the velocity field has the following analytical solution 
\begin{equation*}
    v'(t) = A\sin(kx-\omega t) + A^2\frac{\gamma+1}{4} \frac{\omega t}{c_{s,0}}\cos{[2(kx-\omega t)]}.
\end{equation*}
valid up to second-order in a perturbative sense (hence the $A^2$ dependence). Using $A = 10^{-5}$ allows us to describe the exact solution with only a second-order accurate formula, as higher order corrections will only show up with an amplitude smaller or equal to $A^3=10^{-15}$ below machine precision.

We compare the numerical solution to the analytical one at time $t=\ell/c_{s,0}$, which corresponds to the time it takes for the nonlinear sound wave to perform one complete orbit over the periodic domain. Both solutions are compared using control volume averaged quantities. 

\begin{figure}
    \centering
    \includegraphics[width=\columnwidth]{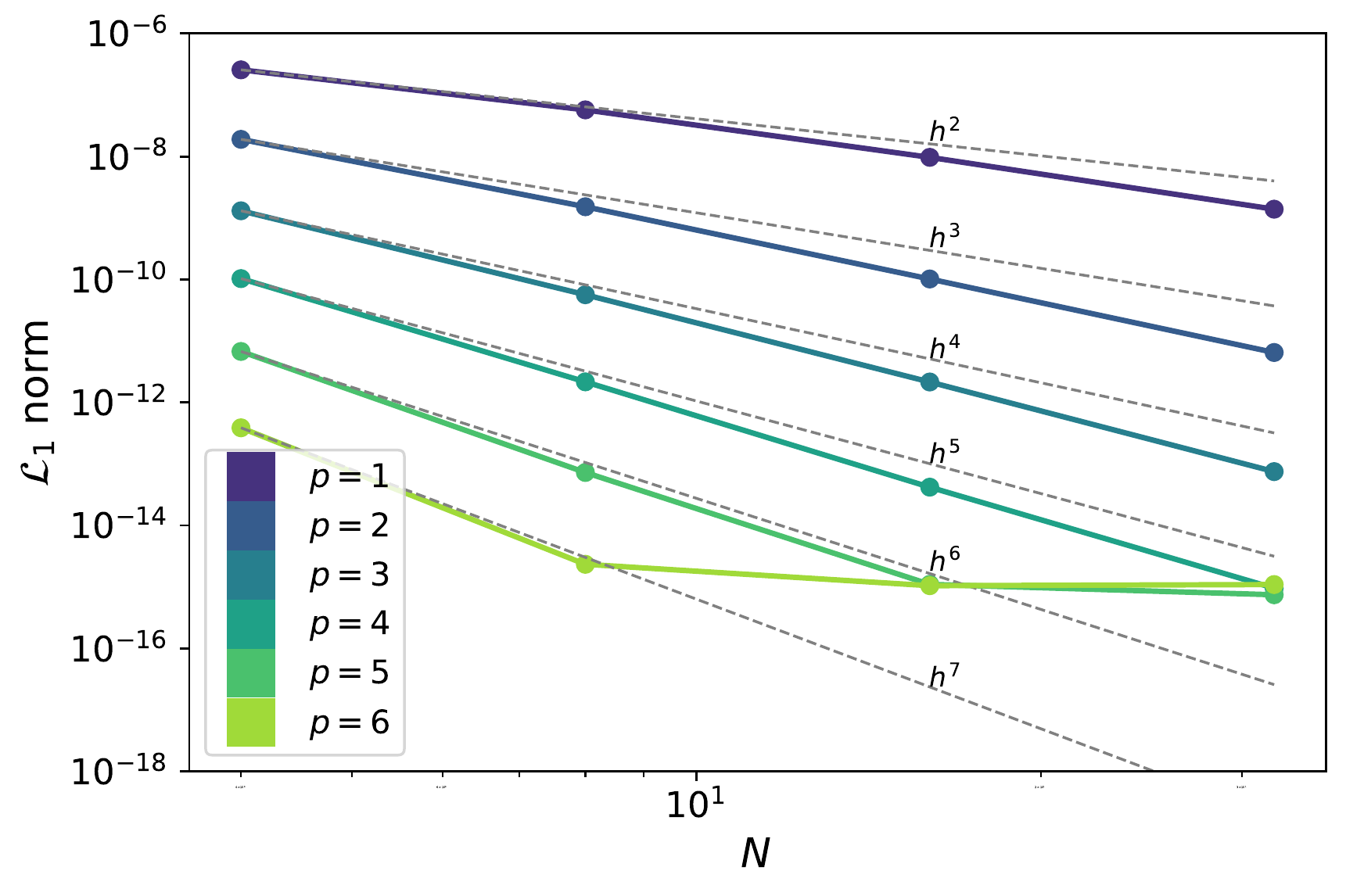}
    \caption{Convergence study using the $\mathcal{L}_1$ norm for the SD method and the nonlinear sound wave test at different orders and spatial resolutions.}
    \label{fig:soundwave}
\end{figure}

Fig.~\ref{fig:soundwave} shows a convergence study for this test, using the error in the $\mathcal{L}_1$ norm, showing that the SD method converges slightly faster than expected for this test. All simulations were performed with the troubled cell detection enabled, in order to check that our algorithm, in particular the smooth extrema detection, is preserving this perfectly smooth solution.

\subsubsection{Kevin-Helmholtz instability}
\label{sect:KH}

The Kevin-Helmholtz instability (KHI) is a phenomenon present in many astrophysical phenomena. This instability is triggered by shear motion. Its nonlinear evolution leads to the formation of large vortices (eddies). These large vortices trigger secondary instabilities at small scales, leading to the onset of a turbulent cascade across a large range of scales. In numerical simulations of the KHI, the minimum eddy size is determined by the spatial resolution. In other words, while in nature kinetic energy is dissipated on the viscous scales, in our simulations, kinetic energy is dissipated by numerical diffusion. It is therefore particularly interesting to see how our new numerical scheme will behave in such a numerical experiment. The initial conditions for this test are as follows:
\begin{align*} 
(&\rho,v_x, P) = 
\begin{cases}
     (2,1/2, 2.5) \quad 0.25<y<0.75 \\
     (1,-1/2, 2.5) \quad \text{otherwise},
\end{cases}\\
&v_y = \omega_0 \sin(4\pi x)\left[e^\frac{-(y-1/4)^2}{(2\sigma^2)} + e^\frac{-(y-3/4)^2}{(2\sigma^2)}\right],\\
\end{align*}
where $\sigma = 0.05\sqrt{2}$ and $\omega_0=0.1$. The computational box has dimension $[0,1]\times[0,1]$ with periodic boundary conditions.
\begin{figure*}
    \centering
    \includegraphics[width=0.95\textwidth]{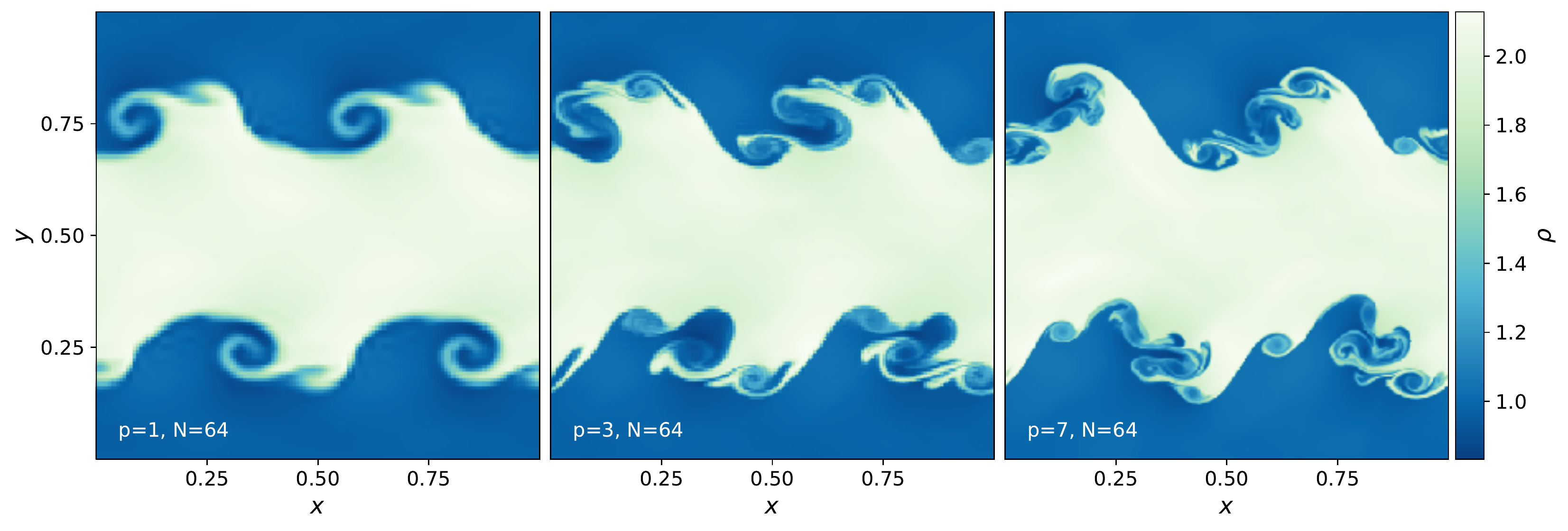}
    \caption{Color maps of the density control volume averages of each subcell for the Kevin-Helmholtz instability test at $t=0.8$. The simulations where performed with $64^2$ cells for $3$ values of the polynomial degree $p=1,3$ and $7$.}
    \label{fig:KH}
\end{figure*}
\autoref{fig:KH} shows the control volume averaged density for each subcell at $t=0.8$ and for different polynomial degrees $p=1,3$ and $7$, on a mesh with $64^2$ elements. This problem is particularly hard for FE methods because of the presence of a discontinuity. Our results show that the small scale structure of the instability is better and better captured as one increases the polynomial degree. This is a highly non trivial result made possible by our a posteriori limiting strategy. 

\begin{figure*}
    \centering
    \includegraphics[width=.9\textwidth]{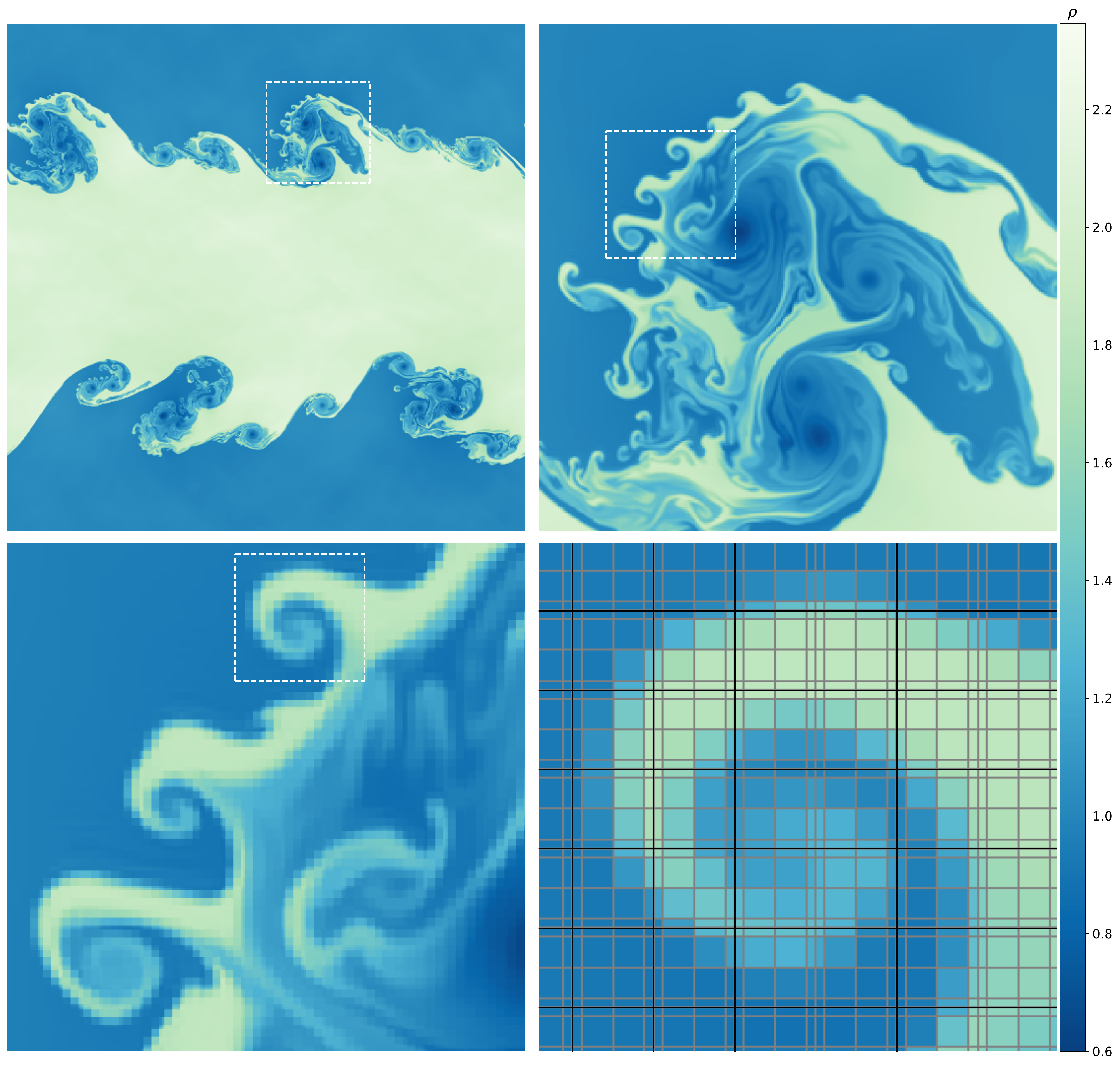}
    \caption{High resolution simulation for the Kevin-Helmhotz instability. Color map of the density control volume averages of each subcell for the Kevin-Helmholtz instability test at $t=0.8$. The simulation was performed with $512^2$ cells and a polynomial degree of $p=3$, totalling $2048^2$ degrees of freedom (per variable).}
    \label{fig:KH_fs}
\end{figure*}

\autoref{fig:KH_fs} shows our highest resolution simulation of the KHI with $512^2$ elements and $p=3$ at $t=0.8$. This corresponds to $2048^2$ numbers of DOF. By progressively zooming on the secondary instabilities, we  show in \autoref{fig:KH_fs} that the sharp discontinuity is captured at the subcell level, well within the elements. When compared to \citet{Schaal2015}, who used adaptive mesh refinement (AMR) and effectively $4096^2$ elements with also $p=3$ but this time using DG, our results look sharper, although we used much less elements ($512$ per dimension here versus $4096$ per dimension there). We attribute this to the better discontinuity capturing capabilities of our a posteriori limiting strategy, compared to their adopted limiting strategy, a common one for the modal DG scheme, which discards higher than $p=1$ order information in the presence of discontinuities.

\subsubsection{Mach 3 wind tunnel with a step}

We would like now to perform a more difficult test that combines strong shocks with various fluid instabilities in multiple dimensions. A classical test in the context of high-order methods is the Mach 3 wind tunnel with a step \citep{woodward1984numerical, Dumbser2014}. 
The initial conditions for this test are as follows:
$(\rho,v_x,v_y,P) =(\gamma,3,0,1)$
where $\gamma=1.4$. The computational box has dimensions  $[0,3]\times[0,1]$ with a hard step in the region $[0.6,3]\times[0,0.2]$. The left and right boundaries have inflow and outflow boundary conditions respectively, while top and bottom boundaries are reflective (including the hard step). We used the LLF Riemann solver for both the SD method and the MUSCL fallback scheme. The tolerance for the NAD criterion was set to the fiducial value $\epsilon=10^{-5}$. These choices are motivated by the presence of strong shocks in the flow.

\begin{figure*}
    \centering
    \includegraphics[width=.9\textwidth]{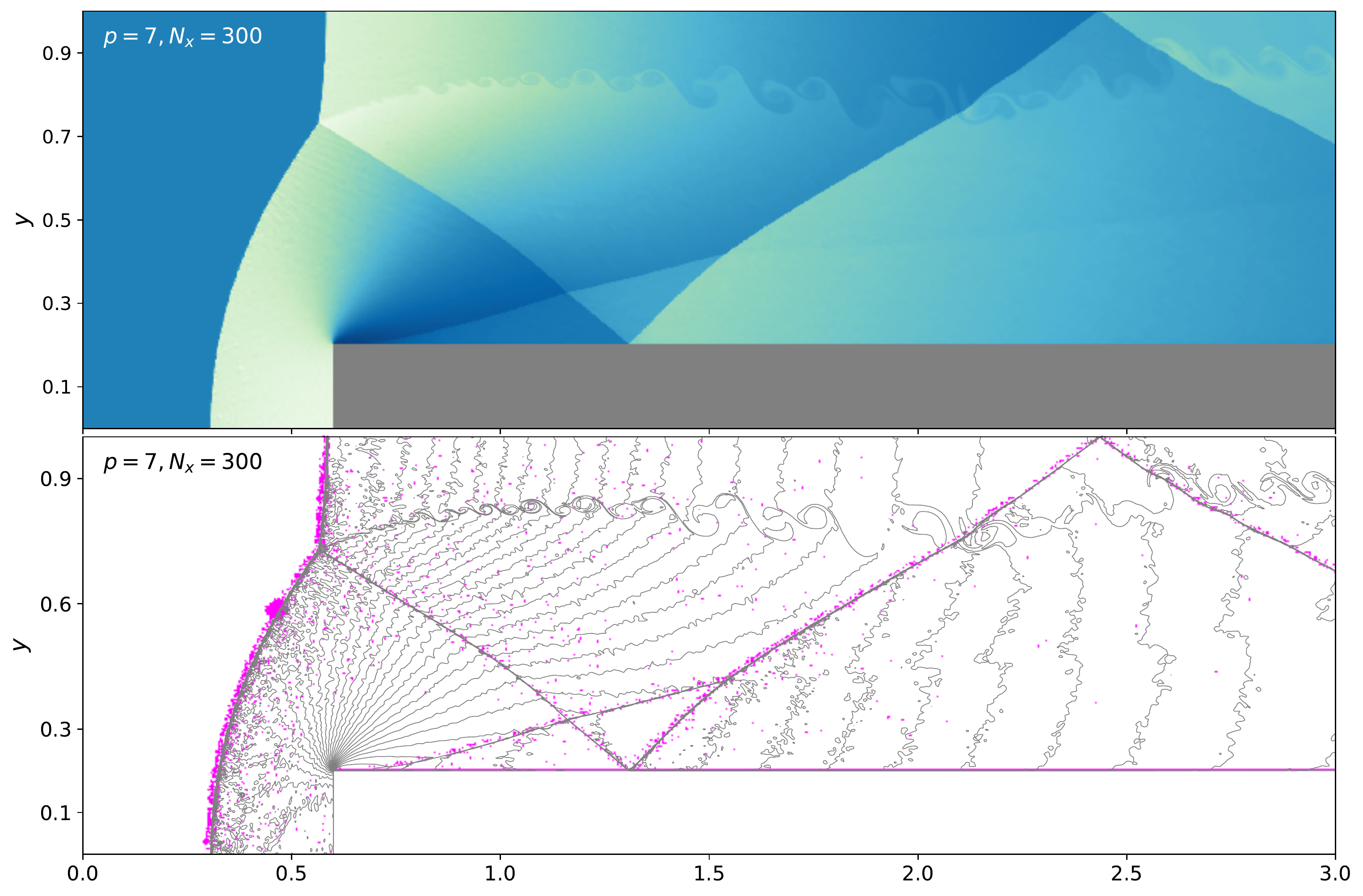}
    \caption{Mach 3 wind tunnel with a step with $N_x = 300$ elements and $p=7$: the top panel shows a map of the density at time $t=4$, while the bottom panel shows a map of the troubled cells (including neighboring cells) superimposed to density isocontours.}
    \label{fig:WT}
\end{figure*}

Figure~\ref{fig:WT} shows our results at time $t=4$ obtained for $N_x=300$ and $p=7$. One can clearly see that shocks and contact discontinuities are properly captured. More interestingly, we also see the train of vortices generated by the triple point about the step. As discussed in \cite{Dumbser2014}, capturing these vortices is a signature of high-order accuracy of our solution. 

The bottom panel shows a map of the troubled cells, together with their immediate neighbors, for which at least one high-order flux was replaced by a second-order one. The density isocontours are useful to visualize that troubled cells are aligned with the main discontinuities of the flow. Some sporadic troubled cells are also visible further away from shocks, maintaining the monotonicity of the solution.

Table~\ref{tab:WT} shows the fraction of the computational volume covered by troubled cells (excluding neighbors). One sees that the fraction of troubled cells steadily decreases with the resolution. Ideally, one expects troubled cells to ultimately cover only the main discontinuities. For a fixed number of DOF, however, the fraction of troubled cells increases slightly with the polynomial degree. This is probably the outcome of the classical Runge phenomenon.

\begin{table}
    \centering
    \begin{tabular}{c|c|c|c}
    \hline 
    $(p+1)N_x=$  &  $600$ & $1200$ & $2400$  \\
    \hline 
      $p=1$   &  $2.02\%$ & $1.23\%$ & $0.8\%$ \\
      $p=3$   &  $2.63\%$ & $1.67\%$ & $1.08\%$ \\
      $p=7$   &  $3.62\%$ & $2.18\%$ & $1.35\%$ \\
    \hline
    \end{tabular}
    \caption{Ratio of average troubled subcells detected per time step at $t=4$ for the wind tunnel test.}
    \label{tab:WT}
\end{table}
\subsubsection{Double Mach reflection}

The second classical test combining strong shocks and fluid instabilities is the Double Mach reflection \citep{woodward1984numerical, Dumbser2014}. Although very similar to the previous one, this test allows the formation of a strong shear flow at the back of the inclined ramp. 

The initial conditions for this test are as follows:
\begin{align*} 
(&\rho,v_x,v_y,P) = 
\begin{cases}
     (8,7.145,-4.125,116.5) \quad &x'<1/6 \\
     (\gamma,0,0,1) \quad &\text{otherwise},
\end{cases}
\end{align*}
where $x' = x - y/\tan(\pi/3)$ and $\gamma=1.4$. The computational box has dimensions $[0,4]\times[0,1]$ with inflow and outflow boundary conditions on the left and right boundaries respectively, the bottom boundary is reflective while the upper boundary is a time-dependent Dirichlet boundary that follows the shock:
\begin{equation}
    x'(t) = \frac{10t}{\sin(\pi/3)} + \frac{1}{6} + \frac{y}{\tan(\pi/3)}
\end{equation}
We also used here the LLF Riemann solver for both the SD and the MUSCL fallback scheme.

\begin{figure*}
    \centering
    \includegraphics[width=.9\textwidth]{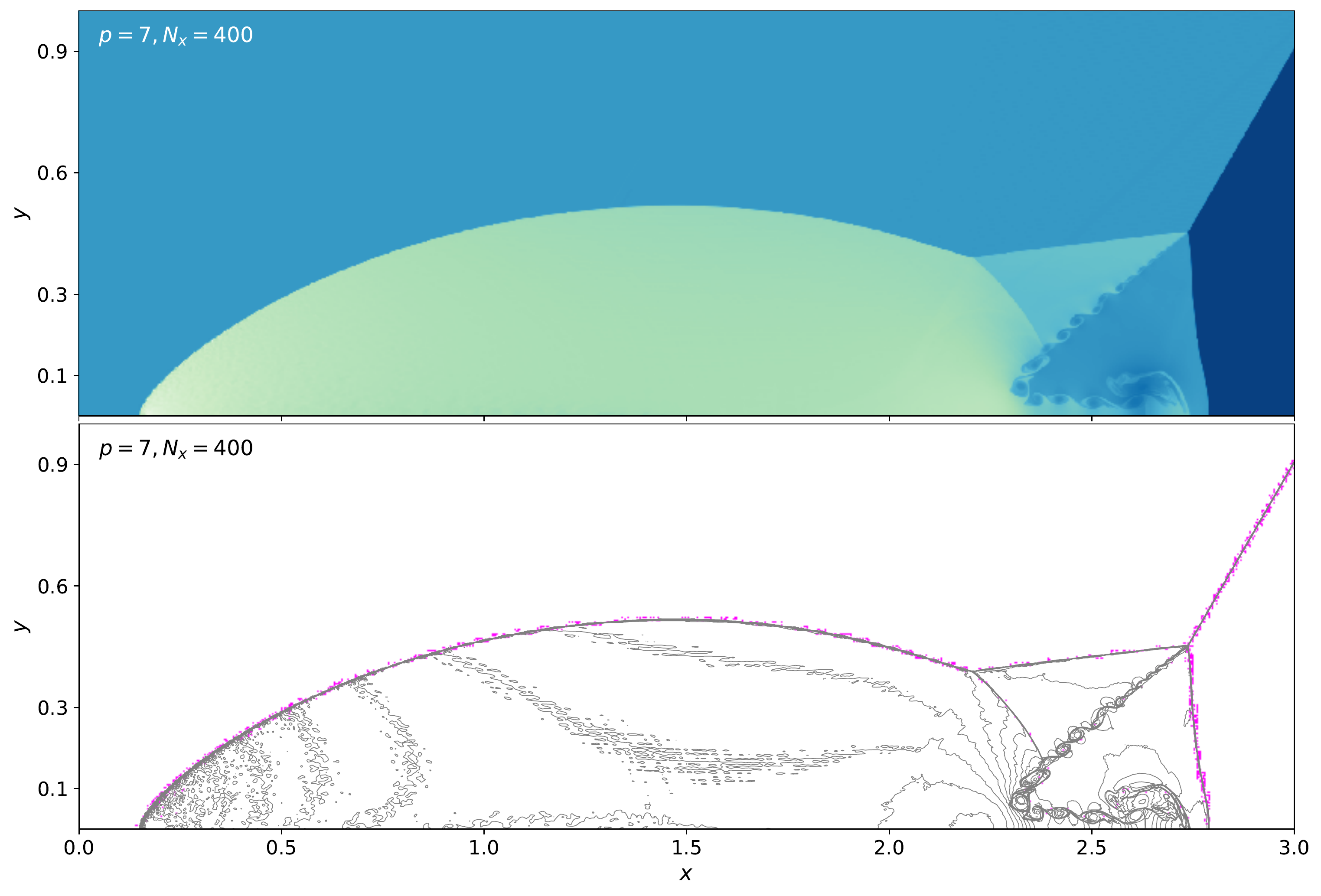}
    \caption{Double Mach reflection with $N_x = 400$ elements and $p=7$: the top panel shows a map of the density at $t=0.2$ while the bottom panel shows the troubled cells (including neighboring cells) superimposed to density isocontours.}
    \label{fig:DM_tr}
\end{figure*}

Figure~\ref{fig:DM_tr} shows both the density map and the location of the troubled cells and their neighbors. The jet that forms on the right side shows the same typical Kelvin-Helmholtz instability as in section~\ref{sect:KH}, except that here it was triggered by interaction of strong shocks. Troubled cells are here again nicely aligned with the main discontinuities. It is interesting to see that the jet region seems to contain very few troubled cells. As a result, the flow exhibits a complex pattern of vortices, as noted also in \cite{Dumbser2014}.

\begin{figure*}
    \centering
    \includegraphics[width=.9\textwidth]{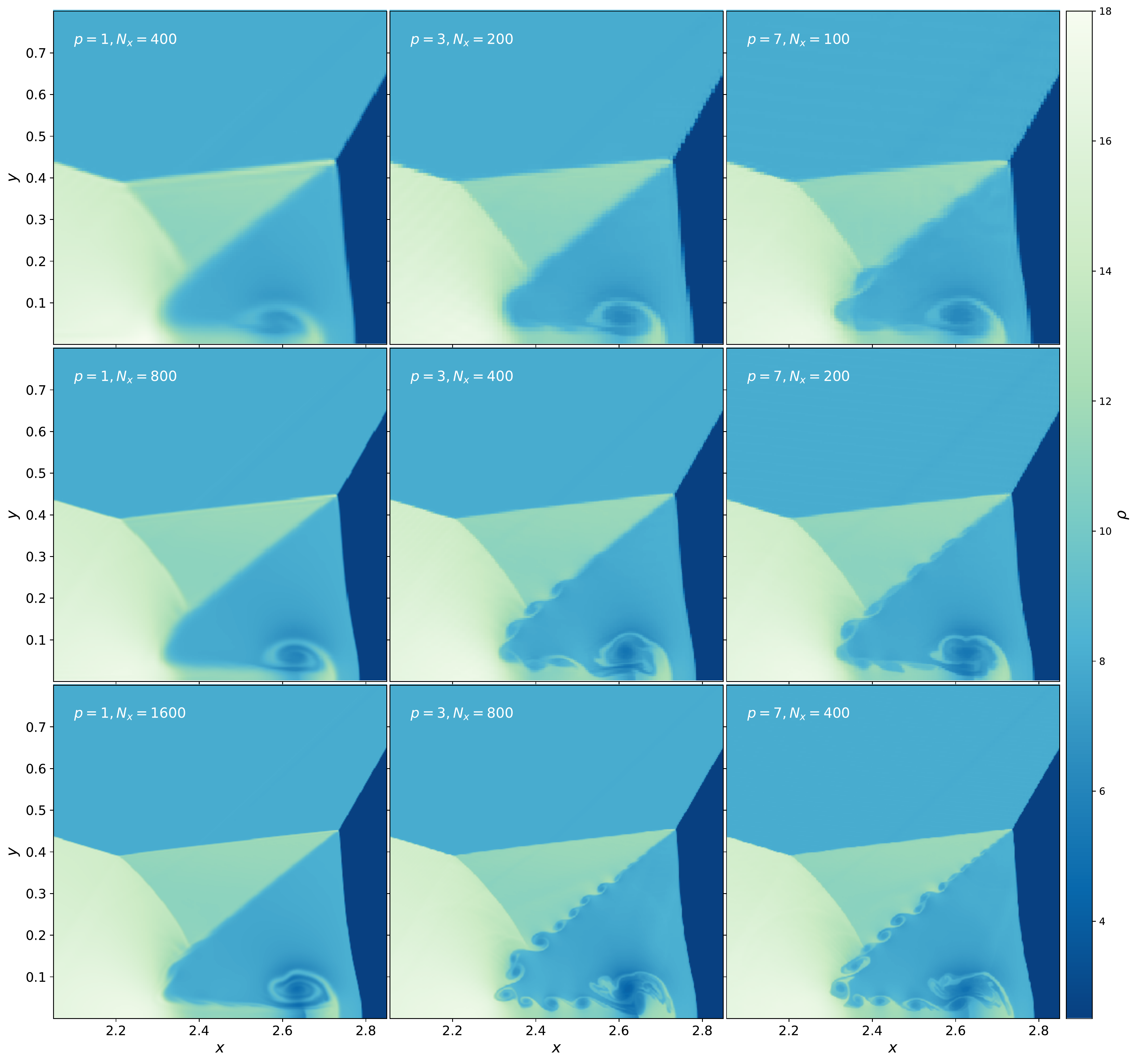}
    \caption{Double Mach reflection with different resolution and polynomial degree. Each row corresponds to a given number of degrees of freedom, with the number doubling each time from top to bottom.}
    \label{fig:DM}
\end{figure*}

Figure~\ref{fig:DM} shows a zoom-in view of the jet, with varying resolutions and polynomial degrees. One clearly sees that even for a fixed number of DOF, going to high-order allows a systematically better description of the turbulent flow. In particular, the case with $p=7$ and $N=100$ seems comparable (or arguably better, even) than $p=1$ and $N=1600$ with 4 times more degrees of freedom per dimension. 

Table~\ref{tab:DM} shows the fraction  of troubled cells as a function of resolution and order. We observe similar trends as for the wind tunnel with a step. 

\begin{table}
    \centering
    \begin{tabular}{c|c|c|c}
    \hline 
    $(p+1)N_x =$  & $800$ & $1600$ & $3200$  \\
    \hline 
      $p=1$   &  $0.98\%$ & $0.53\%$ & $0.28\%$ \\
      $p=3$   &  $1.33\%$ & $0.75\%$ & $0.41\%$ \\
      $p=7$   &  $1.64\%$ & $0.88\%$ & $0.47\%$ \\
    \hline
    \end{tabular}
    \caption{Ratio of average troubled subcells detected per time step at $t=0.2$ for the double Mach reflection test.}
    \label{tab:DM}
\end{table}

\section{Discussion}
\label{sec:discussion}

The results obtained in our various tests consistently support a better performance of our novel SD implementation when compared to classical DG implementations with mode by mode limiting \citep{krivodonova2007limiters,Schaal2015}. This positive result was to be expected: indeed, instead of degrading the order of the solution in entire troubled elements, we are limiting the solution in troubled sub-cells within the elements. As observed in \citet{Dumbser2014,Vilar2019} and in the present work, this allows the proper description of discontinuous solutions within the internal DOF of the elements. In contrast, mode by mode limiting tends to lose the internal structure of the solution within elements when exposed to discontinuities. Moreover, we observe that this difference in quality seems to increasing dramatically when going to very high order ($p>4$). This is a non-trivial result, as mode by mode limiting, as the name suggests, lowers the order of the solution one polynomial degree at a time, slowly cascading down to second or even first order. In the present method, when limiting occurs, the solution falls back directly to a second order solution in each sub-cell of the limited element, thus degrading less local information of the approximate solution. 

When comparing our results to those of \citet{Vilar2019}, we obtain similar results, with a slight advantage for the present work. These two methods are quite similar, both making use of a second order fallback scheme for individual troubled sub-cells. The major differences between our work and \citet{Vilar2019} is their use of DG in conjunction with a RK3 time integrator, as opposed to our use of SD with an ADER time integrator. Their time integration is always of third order, whereas our time integration always matches the order of the spatial reconstruction. This could explain the slightly better performance of our method.  

When comparing our results to \citet{Dumbser2014}, we observe a clear advantage for our competitors. Although both methods are based on a posteriori limiting, their fallback scheme is a third order WENO scheme, with a projection to a $(2p+1)^2$ uniform mesh inside each troubled element. Therefore, not only is their fallback scheme higher order than ours, but it also contains $2p+1$ DOF per dimension on a uniform internal mesh, compared to $p+1$ DOF per dimension on a non-uniform mesh in our case. We therefore attribute their better results to the combination of these three factors: higher order fallback scheme, more internal DOF for the fallback elements and finally a uniform mesh in the fallback elements. 

We also note that the method in \citet{Dumbser2014} results in a hybrid scheme, with a higher algorithmic complexity than for our scheme. The main disadvantage of our FV fallback scheme is that it features a non-uniform mesh inside each SD element. This non-uniform mesh leaves an undesired imprint on the solution, as shown in several of our tests and has a direct impact on the accuracy of the FV fallback scheme.

When comparing to high-order FV methods with a priori limiting, like those of \citet{zhao2017new}, we observe that at $p<4$, our method presents similar or even better results. On the other hand, when going to very high order ($p > 4$), we observe clearly better results with the FV methods. 

We propose two possible explanations for this: 1- In our a posteriori limiting scheme, we fall back directly to a second order scheme, as opposed to the a priori limiting of the high-order solution in the FV methodology, which seems to preserve better the quality of the high-order solution. 2- Our non-uniform mesh hinders the quality of the fallback scheme because of the increased truncation error compared to a uniform mesh.\\

We considered particularly difficult tests featuring complex shock interaction (wind tunnel with a step and double Mach reflection), outlining the capability of our method to capture discontinuities while preserving the high-order solution of the turbulent flow. Our method compares well with respect to the method presented in \citet{Dumbser2014} with a similar number of DOF. Interestingly, our troubled-cell maps resemble closely the ones presented by \citet{krivodonova2007limiters} and \citet{Dumbser2014}. Note that compared to \citet{krivodonova2007limiters}, our limiting strategy allows us to go to much high-order.

In Appendix \ref{sec:tolerance_criteria}, we briefly studied the impact of the parameter $\varepsilon$ (the tolerance factor in the \textit{NAD} criterion) which allows small relative variations without triggering the limiter. Namely, if $\varepsilon=0$, the limiter is strictest, and when $\varepsilon\sim 1$, the fallback scheme is almost never triggered. Our experiments with this parameter show that choosing a larger value, close to $\varepsilon = 10^{-3}$, leads to a much less diffusive scheme, with however some spurious oscillations. All these results demonstrate the large impact of the fallback scheme on the overall quality of the solution. This clearly leaves room for improvement, either by simply using a more accurate fallback scheme, by formulating a laxer \textit{NAD} criterion, or by using a more elaborate fallback mechanism, such as the cascade of fallback schemes proposed by \citet{loubere2014new}.

We would like now to highlight possible advantages of the SD method over high-order FV schemes.
\begin{itemize}
    \item The compact stencil of the SD method prevents communication bottlenecks observed in many high-order FV methods when going to very high order.
    \item The tensor operations required for the high-order interpolation in the SD method  benefit from parallel resources such as graphical processing units (GPU). Indeed, we have observed in our implementation that the extra cost due to the $(p+1)^2$ operations (in 2D) needed for the high-order interpolation can be completely absorbed by saturating the GPU. The extra computational cost due to increasing the order of the scheme (for a constant number of DOF) is only due to the increased number of stages in the time integration. 
    \item For SD, the Riemann solver is only needed at the interface between each element, not at the interface between each subcell. This means we need to solve for only $4(p+1)$ Riemann problems for SD instead of $2(p+2)(p+1)$ Riemann problems for FV.
\end{itemize}

\section{Conclusions}
\label{sec:conclusions}

In this work, we have presented a novel implementation of an arbitrarily high-order SD method, with a second order FV solver as  fallback scheme. We have demonstrated that the resulting method has robust shock-capturing capabilities. We believe it takes the best of both worlds between finite element and finite volume methods.
Indeed, we observe no degradation of the high-order character of the solution in smooth regions of the flow, while at the same time we adequately capture discontinuities (if present) at the subcell level, thanks to our second-order FV fallback scheme. Combining these two opposite requirements is possible, as shown in \autoref{sec:sdisfv}, thanks to the strict equivalence between the SD method and its corresponding FV method at the subcell level.

We have validated our implementation through a battery of 1D and 2D tests, both for the advection equation and for the Euler equations. We have compared our results to several state-of-the-art high-order methods, showing that our scheme, based on subcell a posteriori limiting, is quite competitive.

\section*{Acknowledgements}

The simulations included in this work were executed on the Vesta cluster at the University of Zurich, as well as on computational resources managed and supported by Princeton Research Computing, a consortium of groups including the Princeton Institute for Computational Science and Engineering (PICSciE) and the Office of Information Technology's High Performance Computing Center and Visualization Laboratory at Princeton University. MHV acknowledges funding from the Van Loo Postdoctoral fellowship and from the Michigan Institute for Data Science (MIDAS).

\section*{Data availability statement}
The data underlying this article will be shared on reasonable request to the corresponding author.




\bibliographystyle{mnras}
\bibliography{biblio} 




\appendix

\section{Impact of the Tolerance Parameter}
\begin{figure}
    \centering
    \includegraphics[width=\columnwidth]{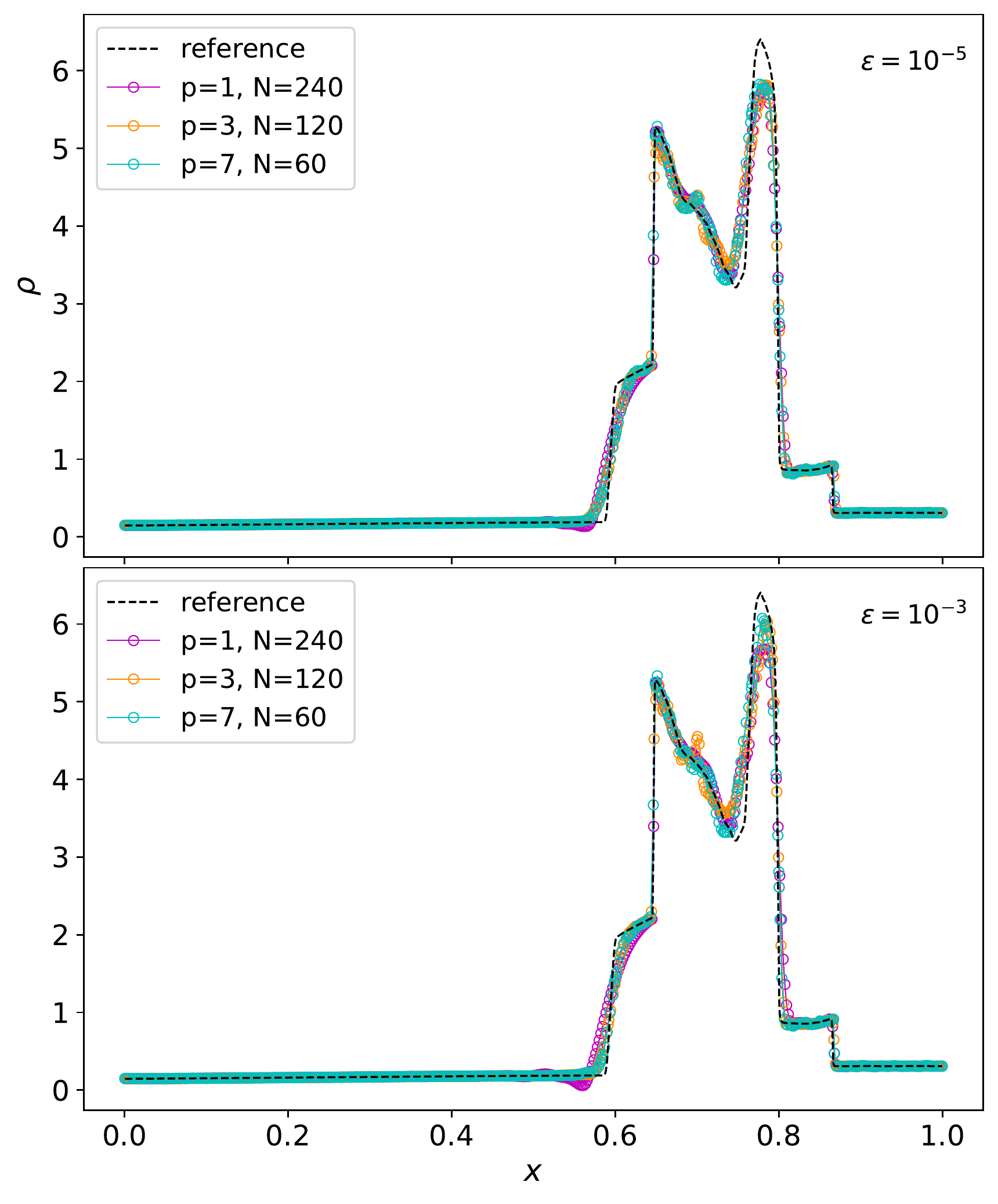}
    \caption{Control volume (subcell) averaged density for the blast wave test of \citet{woodward1984numerical} at $t=0.038$. Top panel for a tolerance factor of $\epsilon=10^{-5}$, while the  bottom panel for a tolerance factor of $\epsilon=10^{-3}$.}
    \label{fig:blast-tolerance}
\end{figure}

In \autoref{fig:blast-tolerance} we present the results for the blast wave problem for two values of the tolerance parameter $\epsilon = 10^{-5}$ and $10^{-3}$, where the former is the value used for the results presented in \autoref{fig:blast}. Here, we wish to point that a less restrictive tolerance results in sharper results, with the improvement being more evident for $p=7$. We also observe that this comes with the prize of having more oscillations.

\label{sec:tolerance_criteria}

\begin{figure}
    \centering
    \includegraphics[width=\columnwidth]{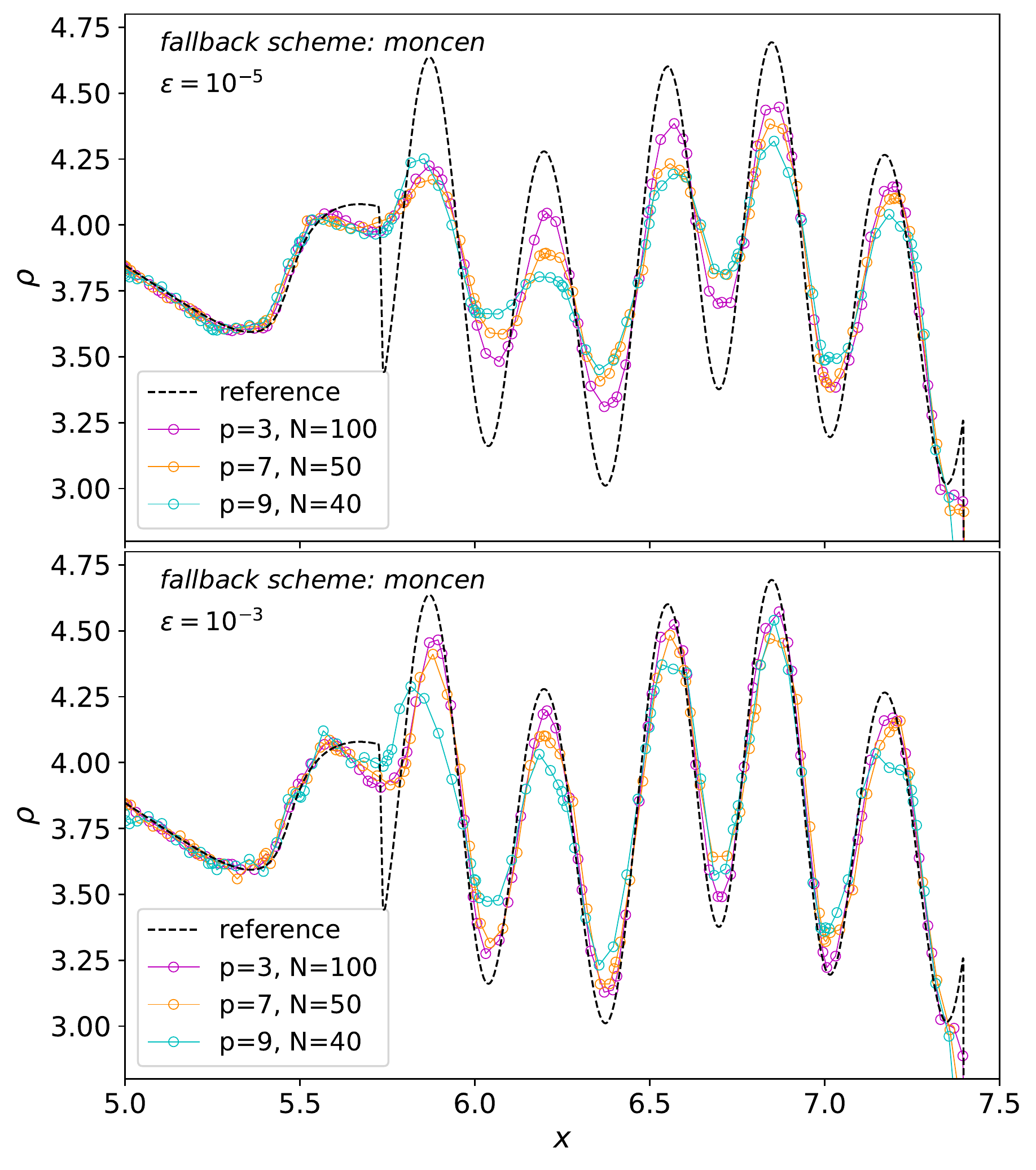}
    \caption{Control volume (subcell) averaged density for the Shu-Osher test \citep{shu1988efficient} at $t=1.8$. Top panel for a tolerance factor of $\epsilon=10^{-5}$, while the bottom panel for a tolerance factor of $\epsilon=10^{-3}$.}
    \label{fig:shu-tolerance}
\end{figure}

In \autoref{fig:shu-tolerance} we present the results for the Shu and Osher problem for the same two values of the tolerance parameter. We show the results for a DOF$=400$, half of that used on the top right panel of \autoref{fig:shu}. It is quite obvious here that a less restrictive tolerance parameter results in a better agreement with the expected solution. In this case, the quality of our solution becomes even comparable to the best published results obtained using FV schemes of similar order of accuracy.

These results suggest that our fiducial limiting parameters are probably too conservative, and therefore, there is room for improvement, be it through a different, more sophisticated tolerance criteria, or through a cascade of fallback schemes of progressively decreasing order.


\bsp	
\label{lastpage}
\end{document}